\documentclass[traditabstract]{aa}

\usepackage{graphicx}
\usepackage{natbib}
\usepackage{xcolor}

\usepackage{txfonts}
\begin{document}

\title{Measurement of interstellar extinction for classical T Tauri stars using far-UV \mbox{H$_{2}$} line fluxes }
%\subtitle{Behaviour of the Paschen lines during flares and quiescence}

\author{B. Fuhrmeister\inst{\ref{inst1},\ref{inst2}}, P.~C. Schneider\inst{\ref{inst2}}
  \and Th. Sperling\inst{\ref{inst1}}
  %\and G. J.  Herczeg\inst{\ref{inst3}, \ref{inst4}, \ref{inst5}}
  \and K. France\inst{\ref{inst6}}
  \and J. Campbell-White\inst{\ref{instESO}}
  \and  J. Eisl\"offel\inst{\ref{inst1}}}

\institute{Th\"uringer Landessternwarte Tautenburg, Sternwarte 5, D-07778 Tautenburg, Germany\\
email{bfuhrmeister@tls-tautenburg.de}\label{inst1} %06
        \and
        Hamburger Sternwarte, Universit\"at Hamburg, Gojenbergsweg 112, 21029 Hamburg, Germany\label{inst2}%01
        \and
%        Kavli Institute for Astronomy and Astrophysics, Peking University, Beijing 100871, Peopleʼs Republic of China\label{inst3}
%        \and
%        Department of Astronomy, Peking University, Beijing 100871, Peopleʼs Republic of China\label{inst4}
%        \and
%        Visiting astronomer, Department of Astronomy; California Institute of Technology; Pasadena, CA 91125, USA\label{inst5}
%        \and
        Laboratory for Atmospheric and Space Physics, University of Colorado Boulder, Boulder, CO 80303, USA\label{inst6}
        \and
        European Southern Observatory, Karl-Schwarzschild-Strasse 2,
85748 Garching bei M\"unchen, Germany\label{instESO}
}

\date{Received dd/mm/2024; accepted dd/mm/2024}

\abstract
    {Understanding the interstellar and potentially circumstellar extinction in the sight lines of classical T Tauri stars is an important ingredient for constructing reliable
    spectral energy distributions, which catalyze protoplanetary disk chemistry, for example. Therefore, some attempts of measuring $A_{V}$ toward individual stars have been made using partly different wavelength regimes and different underlying assumptions. We used strong lines of Ly$\alpha$ fluorescent \mbox{H$_{2}$} and derived the extinction based on the assumption of optically thin transitions.
    %We calculate $A_{V}$ and $R_{V}$ values for TW~Hya, SY~Cha, DK~Tau, and MY~Lup, whose literature $A_{V}$ values span a range between 0. and 1.3\,mag, determined in the optical range and for the canonical $R_{V}=3.1$. We find systematically higher values of $A_{V}$ and for two stars $R_{V}<3.1$, which may hint at small grains in the vicinity of these young stars.
    We investigated a sample of 72 classical T Tauri stars observed with the Hubble Space Telescope in the framework of the ULLYSES program. We computed $A_V$ and $R_V$ values for the 34 objects with sufficient data quality and an additionally $A_V$ value for the canonical $R_V=3.1$ value. Our results agree largely with values obtained from optical data. Moreover, we confirm the degeneracy between $A_{V}$ and $R_{V}$ and present possibilities to break this. Finally, we discuss whether the assumption of optical thin lines is valid.}
    
\titlerunning{Reddening determination by \mbox{H$_{2}$} measurements}
\authorrunning{B. Fuhrmeister et~al.}
\maketitle

%-----------------------------------

\section{Introduction}

Reddening laws predict the visual extinction $A_{V}$=E(B-V)$\cdot$ $R_{V}$
depending on the wavelength, here, the Johnson V and B band. Reddening laws use some
measure of the total dust column E(B-V) \citep{Predehl1995} and a parameter to
characterize the grain distribution $R_{V}$ with a canonical
value of $R_{V}$=3.1 for the interstellar medium that is mainly derived from optical and IR data \citep{Schultz1975, Rieke1985}. Extinction curves show an increase toward the UV, whose slope is characterized by the size of the grains causing the extinction. Large grains produce flatter curves, that is, closer to wavelength independent or gray extinction. These curves are characterized by 
$R_{V}$ values well above 3.1, and measurements up to about $R\sim5.6$ have been reported for dense molecular clouds  \citep{Cardelli1989}.
Rayleigh
scattering would produce steep distributions with $R_{V}$=1.2 (while the steepest
measurement yields about $R_{V}$=2.1; see \citet{Welty1992} and the review of \citet{Draine2003} for further details).
Higher extinction values $A_{V}$ and steeper distributions $R_{V}$ lead to a more
pronounced reddening, that is, a greater loss at bluer wavelengths, especially in the
UV, and they therefore alter the spectral energy distribution
(SED) of an object, by which the UV part is most strongly affected.

Knowledge of the absolute UV radiation is crucial in the context of young cool stars, namely classical T Tauri
stars (CTTSs). In particular, the UV excess is commonly used to derive mass-accretion rates, and it is therefore crucial for the
development of protoplanetary disks (see, e.g.,
\citet{Schindhelm2012} and references therein).

The protoplanetary disks themselves
may contribute to the extinction in addition to the ISM when the line
of sight passes through them. The gas-to-dust ratio of parts of the protoplanetary
disks can be similar to that of the ISM \citep{Schneider2015}, but it may also be much
lower
\citep{Schneider2018}. In the latter case, extinction is caused by regions with
different grain sizes, and individual extinction values are needed for each
star. A
thorough discussion of the different components in the line of sight for CTTSs was given
by \citet{McJunkin2014}.

  We  estimate the
$A_{V}$ values using the far-UV (FUV; $\sim$912-1700\,\AA) spectrum of the star
itself using Hubble space telescope (HST) data covering 1150-1650\,\AA. We specifically used the strongest fluorescent lines of \mbox{H$_{2}$} pumped
by Ly$\alpha$ flux incident on the warm disk.
Cold \mbox{H$_{2}$} ($\approx$ 10\,K) does not radiate effectively because
it has no permanent dipole \citep{Sternberg1988}, and therefore, the observed quadrupole
ro-vibrational transitions in the IR are typically weak. In contrast, %(912-1700 \AA)
dipole-allowed electronic transitions exist in the FUV and result in  strong
\mbox{H$_{2}$} emission lines. These
are fluorescent lines that are mainly photo-excited (pumped) by Ly$\alpha$.
While these lines were first observed for the Sun by \citet{Jordan1977}, the idea of Ly$\alpha$ pumping was developed by \citet{Shull1978}. More than 100 of these lines have been found in CTTSs \citep{Herczeg2002, Herczeg2006}.
They may originate from more than ten upper levels in the $B^{1}\Sigma^{+}_{u}$ electronic
state \citep{Herczeg2006}.
Since the Einstein coefficients for spontaneous emissions are large for these
levels of \mbox{H$_{2}$} \citep[A$_{ul}\approx10^{8}$\,s$^{-1}$, see ][]{Abgrall1993},
they immediately decay into diverse lower levels in the ground state $X^{1}\Sigma^{+}_{g}$,
depending on the branching ratios B$_{ij}$=A$_{ij}$/$\sum{A_{[\nu', J']}}$,
where [$\nu$', J'] is called a progression and is defined by all transitions
from the upper level [$\nu$', J'] (with $\nu$' the vibrational quantum number
and J' the rotational quantum number) to different $\nu$'' and J''
in the ground state. The line identifications (IDs) are
written as ($\nu$'-$\nu$'')R(J'') for J'-J''=-1 and
($\nu$'-$\nu$'')P(J'') for J'-J''=+1.

These FUV \mbox{H$_{2}$} lines were used for $A_{V}$ measurements by
\citet{McJunkin2016}, who included radiative transport simulations. Some radiative transport effects are expected because the same material that absorbs the pumping photons, mostly stellar Ly$\alpha$ photons, also absorbs part of the then  fluorescently emitted H$_2$ photons. This is particularly important because this self-absorption affects the flux ratios of different H$_2$ lines in a qualitatively similar pattern as absorption by dust grains, that is, the observed flux at shorter wavelength is reduced. Therefore, there is some degeneracy in the $A_V$ estimate and in the optical depth effect, which may be addressed using systems with low $A_{V}$.

We attempted to use H$_2$ line flux ratios without any radiative transport calculations, that is, we assumed that the lines are optically thin. This simplified the approach, and we did not need to specify an a priori unknown geometry and density structure. Nevertheless, we are aware that not all \mbox{H$_{2}$} lines may be treated under the assumption that they are optically thin. Our rationale therefore was the following: When we compare the results of progressions that include lines at a short wavelength that may be hampered by self-absorption with the results from progressions that only encompass lines at a longer wavelength where self-absorption becomes negligible, we may be able to identify possible contributions of self-absorption, especially when we also use stars that are known to have low $A_{V}$ values.

Our paper is structured as follows: In Sect. \ref{sec:obs} we briefly describe
the archival data we use and the line flux measurements. Our novel method for obtaining
$A_{V}$ is explained in Sect. \ref{sec:method}. We present our results in
Sect.~\ref{sec:results}, and we discuss the difficulties of the
method in Sect. \ref{sec:discussion}. We conclude in Sect.~\ref{sec:conclusion}.

\section{Observations and line flux measurements}\label{sec:obs}

\subsection{Archival data}

The FUV spectra analyzed here were all part of the data release of the Ultraviolet Legacy Library
of Young Stars as Essential Standards (ULLYSES) program \citep{Ullyses2020},
which dedicated 487 HST orbits to low-mass stars, but also included older data. The spectra were
taken with the Cosmic Origins Spectrograph (COS; \citet{Green2012}) with the
medium-resolution G130M and G160M FUV modes, except for TW~Hya, where we used high-resolution data taken with the Space Telescope Imaging Spectrograph (STIS) in E140M and
E140H mode. We used high-level science products (HLSP) spectra, which have an absolute flux calibration and are provided by the
DR5\footnote{\url{https://ullyses.stsci.edu/ullyses-download.html}}
of ULLYSES. We did not apply further reduction steps. We applied a radial velocity correction only for TW Hya because the resolution is much better for STIS. A library of the strongest
\mbox{H$_{2}$} lines from the progressions we used and of the atomic FUV lines
for all ULLYSES stars was published by \citet{France2023}.

Our starting sample contained all 71 targets from \citet{France2023}, where we excluded stars with low luminosity in their $L_{\rm H2}$ measurements (scaled by stellar distance), which left 51 stars. Since the applied threshold was chosen only to filter out the worst cases, we reduced the sample further to 33 stars by investigating the line signal-to-noise (S/N) ratios as described in Sect.~\ref{h2measurement}. 
Moreover, we wished to approach the optical depth problem, and to do this, we needed stars with a literature $A_{\rm V} = 0.$, only a few of which have a sufficient S/N in the H$_2$ lines. We therefore added TW~Hya to our sample. We list the basic
stellar parameters of all 52 preselected stars (including TW~Hya) with sufficient $L_{\rm H2}$ in Table~\ref{tab:stars}. Additionally, we list the number of lines with a low S/N.

\begin{table}
        \caption{\label{tab:stars} Stellar parameters. }
\footnotesize
\begin{tabular}[h!]{lcccccccccc}
\hline
\hline
\noalign{\smallskip}
star     & M$_{*}$$^{a}$ [M$_{\odot}$]& A$_{\mathrm{V}}$ [mag] & references & S/N$^{b}$\\
\noalign{\smallskip}
\hline
\noalign{\smallskip}
TW Hya &  0.77 & 0.0 & McJ14, Ing13 & 1\\%Ingleby 2013: K7, M=0.8 AV=0.0
DM Tau &  0.50 & 0.0--1.5 & Carv22 & 0\\
RECX 11 & 0.83 & 0.0--0.7 & McJ16, Fran23 & 1\\
RECX 15  & 0.20 & 0.0 & Ing13, Rug18 & 1\\
CVSO 90   & 0.62& 0.0 & Cal05 & 6\\
CVSO 109 & 0.46 & 0.1 & Fran23 & 13\\
Sz 69 & 0.20 & 0.0 & Alc17 & 17\\
Sz 77 & 0.75 & 0.0, 0.3 & Alc17, Fran23 & 3\\
Sz 84 & 0.16 & 0.0 & Alc17 & 10 \\
Sz 97 &  0.23 & 0.0 & Alc17 & 14\\
Sz 99 & 0.23& 0.0 & Alc17 & 14 \\
Sz 100 &0.16 & 0.0 & Alc17 & 8 \\
Sz 104 & 0.16 & 0.0& Alc17 & 16 \\
Sz 110 & 0.22 & 0.0 & Alc17 & 10 \\
Sz 111 &  0.58  & 0.0,  0.85 &  Alc17,  Her14 & 0\\
2MASS J0439$^{c}$ &0.17 & 0.0 & Fran23 & 17\\
V510 Ori & 0.76 & 0.1 & Fran23 & 8\\
\hline
AA Tau & 0.80 & 0.40 -- 1.8 & Carv22 & 2\\
CS Cha & 1.05 & 0.8 & Man17 & 0\\
DE Tau &0.59  & 0.3--0.89 & McJ16 & 2\\
DK Tau A & 0.7 & 0.5--2.0   & Carv22 &2 \\ %Ingleby 2013: AV=1.3
DR Tau &0.80  & 0.45--1.70 & Carv22 & 5 \\
DN Tau &  0.60 & 0.55 & Fran23, Her14 & 8\\
HN Tau & 0.85 & 0.5--1.70 & Carv22 & 0\\
MY Lup & 1.06 & 1.30 & Alc17 & 3\\
RY Lup & 1.71 & 0.4 & Alc17 & 1\\
SY Cha & 0.78 & 0.5 & Man17 & 1\\%name in reference: T4
UX Tau A & 1.30 & 0.2--0.64 & McJ16, Fran23 & 5 \\
XX Cha & 0.29 & 0.3 & Fran23 &2 \\
Sz 19 &  2.08  & 1.5 & Fran23 & 11\\
Sz 45 & 0.56 & 0.7 & Fran23 & 3\\
Sz 66 &   0.29 & 0.5, 1.0 & Her14, Alc17 & 2 \\
Sz 68 &  1.40  & 1.0 & Her14 &  4\\
Sz 71 &  0.37 & 0.5, 0.7 & Alc17, Her14 &  5\\% AV also from Fran23
Sz 72 &  0.37  & 0.75 & Alc17 & 8\\
Sz 75 &  0.82  & 1.0, 1.6 & Fran23, Her14 & 1\\
Sz 98 &  0.70& 1.0, 1.25   & Fran23,Her14 & 1\\
Sz 102 &  0.24  & 0.71, 1.13& Alc17, Fran23 & 1\\
Sz 103 & 0.22 & 0.7, 0.85 & Alc17, Her14 & 0\\
Sz 114 & 0.21 & 0.3 & Alc17 & 1 \\
Sz 129 & 0.79 & 0.90 & Alc17 & 1 \\
CHX 18n & 0.81 & 0.8 & Man17 & 3\\
CVSO 58  & 0.81 & 0.8 & Fran23 & 11\\
CVSO 107 & 0.53 & 0.3 & Fran23 & 10 \\
CVSO 146 & 0.86 & 0.6, 0.37 & Fran23, Cal05 & 4 \\
CVSO 176 & 0.25 & 1.0   & Fran23 & 7\\
LkCa 15 & 0.85 & 0.3--1.1 & Carv22 & 2\\
HN 5     & 0.18 & 1.1   & Fran23     & 11\\
RX J1556.1-3655& 0.5 & 0.6, 1.0    & Her14, Alc17  &3\\
sstc2d1608$^{d}$ & 1.4 & 0.2 & Alc17 & 5 \\
sstc2d1612$^{e}$ & 0.44 & 0.8 & Alc17 & 1\\
V505 Ori & 0.81   & 1.0 & Fran23     &10\\

\noalign{\smallskip}
\hline

\end{tabular}
\tablebib{Alc17: \citet{Alcala2017}, Cal05: \citet{Calvet2005}, Carv22: \citet{Carvalho2022}, Fran23: \citet{France2023}, %Ga22: \citet{Gangi2022},
Her14: \citet{Herczeg2014}, Ing13: \citet{Ingleby2013}, Ken95: \citet{Kenyon1995}, McJ14: \citet{McJunkin2014}, Man17: \citet{Manara2017}, %Nel23: \citet{Nelissen2023}
Rug18: \citet{Rugel2018}}\\
$a$: from \citet{France2023} \\
$b$: number of lines affected by S/N$<$10\\
$c$: 2MASS J04390163+2336029\\
$d$: SSTc2d J160830.7-382827\\ %SSTC2dj160830-382827\\
$e$: SSTc2d J161243.8-381503\\ %SSTC2dj161243-381503\\
\normalsize
\end{table}

\subsection{Literature $A_{V}$ values}

%To compare the H$_2$-based $A_V$-values with 
%the ``correct'' extinction values, we searched the literature for extinction mesurements for the stars of our sample. 

Generally, individual extinction measurements were derived with different methods using
different wavelength regimes. This  may lead to differences in the derived $A_{V}$ values, which may still be found
with the same methods, and even without invoking an intrinsic variability.
%Many methods rely on the comparison of a reddened star with an unreddened template star (or model spectrum)
%in photometric colours or a spectrum, which always comes with the difficulties of finding the correct template or
%model spectrum. 
An overview of the different measurement methods and a comparison of
$A_{V}$ values for
stars in the Taurus region can be found in \citet{Carvalho2022}, and 
a number of well-studied CTTSs was described in \citet{McJunkin2016}. The derived extinction values occasionally vary
substantially. For example, the $A_V$ values (assuming $R_{V}$=3.1) for DK~Tau~A  range from
0.46\,mag \citep[][derived from Ly$\alpha$ reconstructions]{McJunkin2014} to 1.99\,mag
\citep[][derived from IR data]{Luhman2017}; the latter was converted from A$_{J}$ with
the relation $A_{V}$/A$_{J}$=3.55 using the extinction curve by \citet{Mathis1990}.
Typical variations in $A_V$ measurements are about $\sigma(A_V)\sim 0.6$\,
mag \citep{Grankin2017, Carvalho2022}. Nevertheless, many literature $A_V$ values for \object{DK~Tau~A} fall in
the range between $A_{V}$=0.8 and 1.4\,mag \citep{Carvalho2022}, even though they
were obtained with different methods at different wavelengths.

%In the case, when values obtained from optical or infrared measurements are available, we concentrate on values derived from optical data, for example, for the stars
%MY~Lup and SY~Cha. For both stars, \citet{Manara2017} and
%\citet{Alcala2017} used broadband VLT/X-SHOOTER spectra to fit the stellar parameters,
%extinction, and mass loss rates simultaneously.

Consistent values of $A_V$  are mainly obtained for $A_{V}\sim0.0$\,mag, for example, for TW~Hya
\citep{Rucinski1983, Ingleby2013, McJunkin2014}. 
The reported values for CVSO~90 and RECX~15 are also zero or are very low \citep{Calvet2005, Ingleby2013}. On the other hand, the range of measured $A_{V}$ values for DM Tau is large:
Although many studies favored a low value of 0.0 -- 0.1\,mag \citep{Kenyon1995, Herczeg2014},
there are also studies that reported much higher values of about 0.7 -- 0.9\,mag \citep{McJunkin2016, Carvalho2022},
especially using IR diagnostics. We list the literature
 $A_{V}$ values for all stars in Table~\ref{tab:stars}.

For some stars mainly from the Taurus star-forming region, we are aware of a number of extinction measurements. For the comparison to our $A_V$ values in Sect.~\ref{sec:discussion}, we therefore used the median of the values listed in \citet{McJunkin2016, Grankin2017}, and \citet{Carvalho2022} after omitting outliers with $A_V > 3.0$\,mag. For all other stars, we are aware of only one or two measurements as listed in Table~\ref{tab:stars}. There, we used the mean, and for stars with only one measurement, we adopted an error of 0.1\,mag (which is most probably too low, considering the spread for stars with many measurements).

\begin{table}
        \caption{\label{tab:lines} Selected \mbox{H$_{2}$}\, emission line properties. }
\footnotesize
\begin{tabular}[h!]{lcccccl}
\hline
\hline
\noalign{\smallskip}

 progression   & line ID$^{a}$  & wave- & A$_{ul}$  & $\tau_{lu}^{b}$& comm.$^{c}$  \\
 + pumping   &          & length& [10$^{8}$\\
 transition   &     &   [\AA] &  s$^{-1}$]& \\
\noalign{\smallskip}
\hline
\noalign{\smallskip}
[1,7]                    & (1-6)R(6) & 1442.87 & 0.9 & $<10^{-3}$\\%$<10^{-5}$&$<10^{-4}$\\
 (1-2)R(6)               & (1-6)P(8) & 1467.08 & 1.3 & $<10^{-4}$\\%$<10^{-5}$&$<10^{-4}$\\
 $\lambda$1215.72\AA     & (1-7)R(6) & 1500.45 & 1.7 &$<10^{-4}$\\%$<10^{-6}$&$<10^{-4}$\\
      & (1-7)P(8) & 1524.65 & 1.9 &$<10^{-4}$\\%$<10^{-6}$&$<10^{-4}$\\
      & (1-8)R(6) & 1556.87 & 1.3 & $<10^{-5}$\\%$<10^{-7}$&$<10^{-5}$&wl\\
      & (1-8)P(8) & 1580.67 & 1.1 & $<10^{-5}$\\%$<10^{-7}$&$<10^{-5}$&wl \\%blend with (2-8)P(12) at 1580.71\AA\\
\hline
[1,4] & (1-6)R(3) & 1431.01 & 1.0 &$<10^{-3}$\\%$<10^{-4}$&$<10^{-3}$\\
(1-2)P(5)      & (1-6)P(5) & 1446.12 & 1.4 & $<10^{-3}$\\%1.4 &$<10^{-4}$&$<10^{-3}$\\
 $\lambda$1216.07\AA      & (1-7)R(3) & 1489.57 & 1.6 &$<10^{-3}$\\%$<10^{-5}$&$<10^{-3}$\\
      & (1-7)P(5) & 1504.76 & 2.0 &$<10^{-3}$\\%$<10^{-5}$&$<10^{-4}$\\
      & (1-8)R(3) & 1547.34 & 1.1 & $<10^{-4}$& bw \ion{C}{iv}\\%$<10^{-6}$&$<10^{-4}$&bw \ion{C}{iv}\\
\hline
[0,1] & (0-4)P(2) & 1338.56 & 3.1 &0.5\\%0.2&0.3\\
 (0-2)R(0)     & (0-5)P(2) & 1398.95 & 2.6 &0.04\\%0.01&0.03\\
 $\lambda$1217.20\AA      & (0-6)P(2) & 1460.17 & 1.5 &$<10^{-2}$\\%$<10^{-3}$&$<10^{-3}$\\
      & (0-7)P(2) & 1521.59 & 0.6 & $<10^{-3}$ & wl\\%$<10^{-4}$& $<10^{-4}$&wl\\
\hline
[0,2] & (0-4)P(3) & 1342.26 & 2.8 &1.6\\%0.8&1.0\\
(0-2)R(1)      & (0-5)R(1) & 1393.96 & 1.6 & 0.15& bw H$_{2}$ \\
 $\lambda$1217.64\AA      & (0-5)P(3) & 1402.65 & 2.3 & 0.16 &bw \ion{Si}{iv} \\%at 1402.77\AA\\
      & (0-6)P(3) & 1463.83 & 1.4 & 0.01\\
      & (0-7)P(3) & 1525.15 & 0.5 & $<10^{-2}$&wl\\
%      &           &         &     & bw H$_{2}$\\%blend with (0-7)R(5) at 1525.23\AA\\
\noalign{\smallskip}
\hline

\end{tabular}
\tablebib{All wavelengths and A$_{ul}$ values are taken from \citet{Abgrall1993}.}\\
$^{a}$ The line notation is as follows: A progression [$\nu$',J'] consists of all transitions from the upper level with a rotational quantum number J' and a vibrational quantum number $\nu$' in the electronic state $B^{1}\Sigma^{+}_{u}$ to all levels with J'' and $\nu$'' in the ground state $X^{1}\Sigma^{+}_{g}$. Thus, the line identifications are written as ($\nu$'-$\nu$'')R(J'') for J'-J''=-1 and
($\nu$'-$\nu$'')P(J'') for J'-J''=+1.\\
$^{b}$ optical depth $\tau_{lu}$  refers to gas temperatures of 2000\,K; the assumptions we made for the calculation are detailed in Sect.~\ref{sec:selfabsorption}.\\
$^{c}$ abbreviations used: wl: weak line; bw: blend with\\
\normalsize
\end{table}

\subsection{Measurement of the \mbox{H$_{2}$} line flux}\label{h2measurement}

For our estimate of $A_V$, we used the four strongest
progressions from the Lyman band, with ro-vibrational state upper levels
[$\nu$', J'] = [1,7], [1,4], [0,1], and [0,2], and from these
all strong  \mbox{H$_{2}$} lines that were also selected by
\citet{Hoadley2015}. We list the progression, line identification, Einstein coefficient
A$_{ul}$, and wavelength
together with some comments on their usability for our purposes
in Table~\ref{tab:lines}.

For our study, slit losses are not expected to play a role. For typical centering accuracies, they should be
well below 10\%, and the relative slit losses between 1300 and 1600\,\AA\, are expected to be even much
lower. Moreover, the binarity of DK~Tau~A is probably not a problem because it is
a wide binary with a separation of about 2.4 arcsec. 

The line shapes may pose a larger problem here.
Since some of the stars exhibit \mbox{H$_{2}$} lines with broad wings, which may originate
in disk winds (i.e., in a location different from the main emission; \citet{Gangi2023}), we decided to
cut them by fitting the lines with a Gaussian. We are aware that a Voigt profile resembles the instrument profile better. To account for this, we exemplarily fit all lines of MY~Lup also with a Voigt profile and found fluxes for all lines that were higher by approximately 20\%, which altered our resulting $A_V$ and $R_V$ values for this star only within our errors since the higher flux values cancel out in the ratios. We therefore decided for the Gaussian fit, since it is more stable, and we assumed that the flux measured by
this Gaussian originates in the inner disk.%, since even for transitional disks, where the
%\mbox{H$_{2}$} emission is created further away from the star, \citet{Hoadley2015}
%found, that the \mbox{H$_{2}$} emission originates well within 10\,AU.
%The line shapes should be
%dominated therefore by Kepler rotation, which can be approximated by a Gaussian.
To improve the fitting in regions with many lines, we determined the mean
background continuum in an adjacent wavelength region for each of our lines and used
this as the offset in the fit. To obtain errors for these
fitted line fluxes, we applied a bootstrap analysis and varied the spectral flux
within its error. We computed the Gaussian flux of the line  1,000 times and took the
standard deviation of these computations as the error of the line flux.
The typical line flux errors are between 2 and 8\% for many stars; the flux errors of very few lines
of these stars lie above 10\%. Nevertheless, the data for a number of stars are noisier. We therefore excluded stars from our analysis whose S/N ratio of the lines was below 10 for more than six lines because the lines then typically become useless. This reduced our sample from 52 to 34 stars.

In addition to these purely statistical errors,
systematic errors in the line fluxes might be caused by flux variability during
the exposure time. From the raw data of TW~Hya, which exhibits the highest observed fluxes of
our sample stars, we estimated these errors to be about 5\% or lower, which is about
the same as we found for the statistical error. We therefore only applied our statistical
errors in our analysis. 

We show a representative collection of our Gaussian fits for the four stars TW~Hya, SY~Cha, DK~Tau, and MY~Lup in figures published on Zenodo, see Sect.~\ref{data-avail}.
%These figures show that the accuracy of the measured fluxes may suffer from line blends.
We comment on the lines with typically unsatisfactory fits in Table~\ref{tab:lines}.
There are two lines with frequently uncertain fluxes: 
The (0-5)P(3) line at 1402.65\,\AA, which is blended with a \ion{Si}{iv} line,
and the (0-7)P(3) line at 1525.15\,\AA, which is a weak line and is blended with another \mbox{H$_{2}$}
line. Both lines are from the [0,2] progression and cannot be used for all or most of the stars in our sample, respectively. We therefore decided to omit these lines
from the further analysis, which left the [0,2] progression with only three lines.
All other lines were only omitted in the analysis
for the stars when they were identified to be problematic.

\section{Computation of $A_{\mathrm{V}}$}\label{sec:method}

To measure $A_{V}$, we relied on an extinction law, that is, on the variation
of the extinction with the wavelength.
There are different families of typically polynomial
parameterizations for this wavelength dependence, out of which those by \citet{Cardelli1989} and \citet{Fitzpatrick1999} are often used in the context of young stars.
We used the \texttt{unred} function of
the Python package PyAstronomy \citep{Czesla2019pya} to (de-) redden the spectra. This function uses the parameterization of the normalized extinction curve by \citet{Fitzpatrick1999} and allowed us to specify E(B-V) and R$_{\mathrm{V}}$. %\equiv $A$_{V}/$E(B-V).
With only this reddening law, the measured line fluxes from the four progressions
[1,7], [1,4], [0,1], and [0,2], their known transition probabilities $A_{ij}$, and the assumption that the lines are optically thin, we used the following method for deriving $A_{V}$. An alternative method is described in Appendix \ref{method1}. For reference, we call these methods Method I and II where needed for clarity.

\subsection{$A_{V}$ from flux ratios (Method I)}

To retrieve a reliable estimate of A$_{\mathrm{V}}$, we used 
the observed flux ratios for each progression and compared them with the theoretically expected value, that is, we considered permutations between the UV-H$_2$ lines and calculated their ratios. For each progression, we first computed the theoretically expected flux ratios from their Einstein coefficients
A$_{ik}$ and their central wavelength by using
\begin{equation}
r_{theo}=\frac{F_{ik}}{F_{im}}=\frac{E_{ik}\cdot n_{i} \cdot A_{ik}}{E_{im} \cdot n_{i} \cdot A_{im}}=\frac{\lambda_{im}\cdot A_{ik}}{\lambda_{ik}\cdot A_{im}},
\end{equation}
where $E_{ik}$ is the level energy, n$_{i}$ is the level population number of the upper level, which is shared by all lines from a single progression, and $\lambda_{ik}$ is the central wavelength of the line.
This only holds for optically thin line transitions (see also \citet{Ardila2002}).
The method was introduced for [S II] lines, which share the same upper level, by \citet{Miller1968}.

We decided to use all permutations of the reliably measured H$_2$, in a similar fashion as
\citet{Hartigan2007}, that is, we did not use only line ratios relative to some
specific line, for example, the reddest line in each progression.
This resulted in $(n(n-1)\cdot 2)$ pairs,
with $n$ being the number of measured line fluxes in the progression
(because we did not consider inverse pairs).

To obtain an estimate of $A_{V}$, we compared the ratios of our dereddened fluxes to
these theoretical ratios. To this end, we calculated dereddened fluxes
on an arbitrary grid
of $R_{V}$ ranging from 1.5 to 6.2 in steps of 0.08 and of $A_{V}$ ranging from 0.01
to 4.44\,mag in steps of 0.075\,mag.
Next, we computed the respective dereddened line ratios and compared
them to the theoretical ratios for each $R_{V}$ and $A_{V}$ combination. Specifically, 
we determined the best-fitting R$_{\mathrm{V}}$ and $A_{V}$ by finding the minimum of the quadratic form
\begin{equation}
C=\sum_{k=1}^{p} \frac{(r_{meas}-r_{theo})^{2}}{\sigma_{k}^{2}},
\end{equation}
where $\sigma_{k}$ is the observational uncertainty of the measured and dereddened
flux ratio $r_{meas}$. Using a logarithmic form of our $C$ variant, \citet{Hartigan2007} showed $C_{log}$ to have similar statistical
properties as $\chi^{2}$. \citet{Hartigan2007} also showed the better use of the information content when all possible ratios are used, even though they are no longer independent of each other. Errors on the dereddened line ratios, $\sigma_{k}$, were obtained 
by error propagation.

We tested the method by calculating some simulations, that is, we constructed a flux vector for
each progression based on the branching ratios and reddened this flux
with a known $A_{V}$ and $R_{V}$. We then added Gaussian noise to the flux vector
according to different relative errors. We used
$A_{V}$=0.8\,mag and $R_{V}$=3.1 and added noise to the simulated fluxes corresponding to a relative
error of 0.05. This is expected to represent most of our data well. We also performed test calculations with an error of 0.08 and found similar results. 

\begin{figure*}
\begin{center}
  \includegraphics[width=0.95\textwidth, clip]{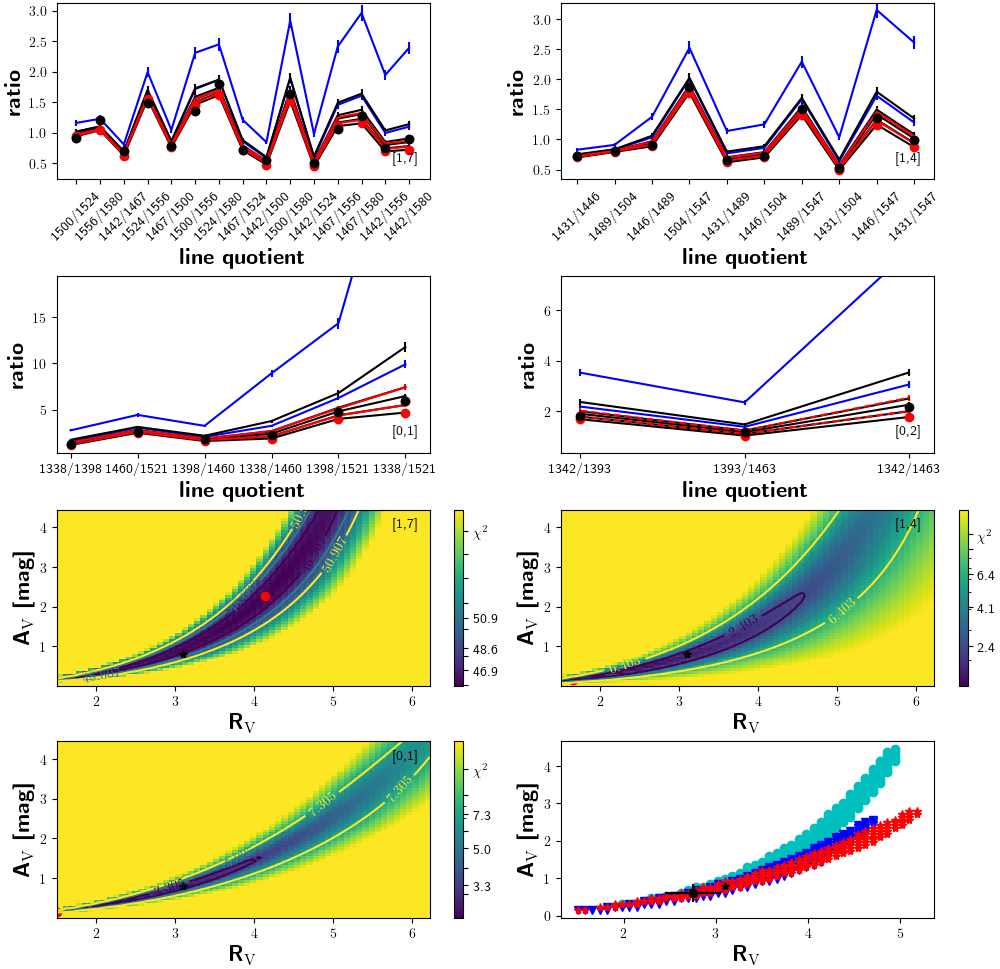}%\vspa4ce{-0.4mm}\\
  \caption{\label{testsim1} Flux ratios for the individual progressions of H$_{2}$
    lines for simulated noisy data with relative flux errors of 0.05. We show the
    progressions as indicated at the lower or upper right corner of each panel. The line quotient
    gives the wavelength of the two involved lines for each ratio. They are ordered by
    increasing wavelength difference between the two lines.
    In the four upper panels, the black dots show the theoretical flux ratios. The red dots
    show the flux ratios from the simulated line fluxes. The black lines
  show the ratios of the dereddened flux (together with the dereddened flux error ratios)
  for $R_V$=3.1 and $A_V$=0.0,0.5,1.0,1.5,3.0. The red and blue lines denote the same for $R_V$=2.1 and 4.1 for $A_V$=1.0 and 3.0, respectively. We note that some of
  the lines are
  almost identical, which again indicating the degeneracy of $R_{V}$ and $A_{V}$.
  %The black line shows the best fit model for each progression.
  Three of the four lower panels show
  contour plots for an increasingly worse fitting of the $R_{V}$ and $A_{V}$ values.
  We show the 68\%, 90\%, and 95\% confidence interval as a black, blue, and yellow line,
  respectively. The numbers in these lines state the $C\sim\chi^{2}$s for the respective
  confidence level and are also given in the color bar. The red dot marks the position of the minimum of the distribution, and
  the black asterisks mark the values chosen
  for the simulation. In the bottom right panel, we overplot the 160 best-fit $C$s
  for every progression. The progression [1,7] is represented by cyan dots,
  progression [1,4] is represented by blue triangles, and progression [0,1] is represented
  by red asterisks. Our retrieved extinction values are marked as a black dot with error bars.%The injected $R_{V}$ and $A_{V}$ combination is again marked by a black asterisk
  %and is located roughly at the intersection of the three progressions.
}
\end{center}
\end{figure*}

\begin{figure}
\begin{center}
   \includegraphics[width=0.5\textwidth, clip]{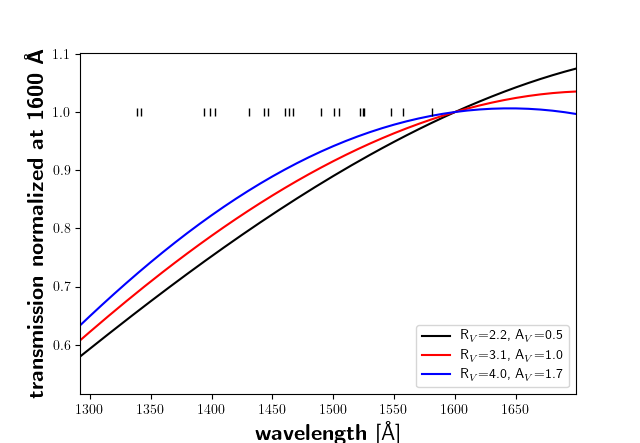}\\%
  \includegraphics[width=0.5\textwidth, clip]{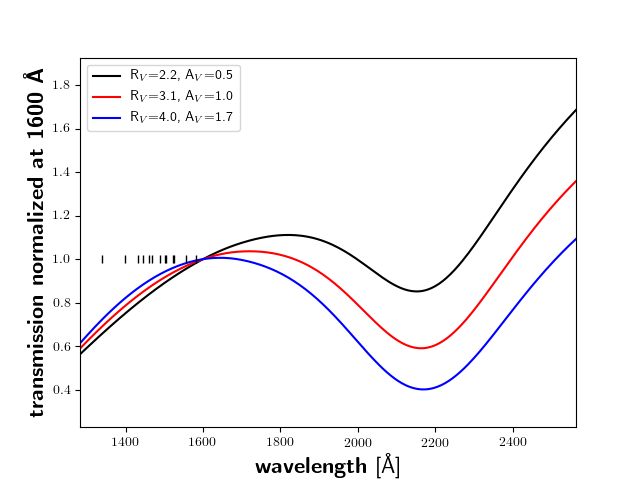}%\vspa4ce{-0.4mm}\\
  \caption{\label{degeneration} Transmission calculated for $A_{V}$ and $R_{V}$
    pairs as given in the legend and normalized at 1600\,\AA, which marks the longest
    wavelength of the lines we considered. The small vertical bars mark the position of the
    lines we used. \emph{Top:} Our considered wavelength range. \emph{Bottom:} Zoom out to show that lines at longer wavelength would probably allow us to break the degeneracy.
}
\end{center}
\end{figure}

We then injected the simulated data in our algorithm. The retrieved
values are shown in Fig. \ref{testsim1}.
For each progression, a degeneracy of $A_{V}$ and $R_{V}$ is visible in the lower half of the panels in Fig.~\ref{testsim1}:
The lower and higher values of $A_{V}$ and $R_{V}$ have rather similar transmission curves
in the considered wavelength range, as was described by
\citet{McJunkin2016}. We show in Fig.~\ref{degeneration}  the transmission curves for three pairs of $A_{V}$ and $R_{V}$. While the shape of the transmission curve is quite similar in the wavelength range covered by UV~H$_2$ lines, which would therefore require data with a high S/N to distinguish between them, the lines at redder wavelength would most probably allow us to break the degeneracy. Unfortunately, there are no H$_2$ lines redward of 1600\AA\, with a sufficiently high S/N for our purposes. \citet{Herczeg2002} detected H$_{2}$ lines up to about 1640\AA  in TW~Hya.
A possible solution may be CO lines, but it is beyond the scope of this paper to investigate these lines with respect to extinction.

An interesting feature shown in the upper four panels in Fig.~\ref{testsim1} is
the different range of ratios
for the different progressions. %While the ratios in the progressions [1,7] and [1,4]
%range from about 0.5 to 1.8, the ratios in the progression [0,1] range from 1.2 to 5.8
%and in the progression [0,2] from 1.2 to 2.2, making 
This shows that progression [0,2] is the least
and progression [0,1] is the most sensitive progression.
The four lower panels in Fig.~\ref{testsim1} show that for all progressions, the derived best-fitting values are found within the 68\% confidence level, although the minimum of the distribution is located at different
positions. Progression [0,2] does not work well for the observed stars, but typically
shows deviating results. Moreover, for many stars, for example, for the stars DK~Tau~A and SY~Cha, it has only two measured line
fluxes and therefore cannot be used for these stars. %(see line fits in Figs.~\ref{dktau1} and \ref{sycha1}. 
%For all other stars it has only three line fluxes to rely on. 
Moreover, it has generally the lowest range of ratios and should therefore be the least sensitive progression. We therefore omitted it from the further analysis. We do not show
the $C$ distribution for this progression here, but replace it by an overlay of the 160 best
$C$ values for each of the three other progressions. The injected $A_{V}$ and
$R_{V}$ value is found roughly at the intersection of the three progressions.
By combining information from
these three progressions, which are by chance influenced differently by the inserted
random noise, the degeneracy between $A_V$ and $R_V$ can therefore be broken, and the true value can be retrieved.

Each progression may best be described by a different combination of $R_{V}$ and $A_{V}$. To obtain the overall best-fit $A_V$ and $R_V$, the $C\sim\chi^2$ of the different progressions is frequently added to find the best-fit values, but this poses several problems in our case. First, the absolute $C$ values differ from one progression to the next, so that one progression may dominate the result of the addition. Second, some progressions have very steep $C$ valleys for low $R_{V}$ and $A_{V}$ combinations, which do not align very well between the different progressions. Furthermore, the $C$ contours fan out for combinations of higher $R_{V}$ and $A_{V}$ values, so that the overall lowest summed $C$ may be found for high $R_{V}$ and $A_{V}$ values. This probably is an effect of the degeneracy between $R_{V}$ and $A_{V}$ and is not real. Therefore, we rather searched for a region in which many good fitting models from different progressions are found. To do this, we used the overlap-region 
of the 160 best-fitting $R_{V}$ and $A_{V}$ combinations for each of the
progressions. This fixed number of best-fitting models approximately corresponds to the 95\% confidence level for many progressions and stars and limits the influence of progressions that contain many more models within the 95\% confidence region.  To identify the position of the largest overlap, we used a box as the search region and counted the number of best-fitting $C$s in this region.
Then we shifted the box for half its size along each axis. The central position for which we found most of the
best-fitting $C$ values in the box was marked as the overall best fit when at least one
$C$ value from each progression was contained in it. When the best-fit $C$
distributions from the different progressions were distinct from each other, the box was enlarged to meet this
additional criterion. The extent of the search box can serve as an error estimate. For the simulations, we used an error box of 0.25\,mag for $A_V$ and a box of 0.3 mag for $R_V$. The derived values are $A_V$=0.6\,mag and $R_V$=2.75, which are lower than the injected values, but agree well within the errors.

\subsection{Possible problems with self-absorption}\label{sec:selfabsorption}
The determination of $A_{V}$ from the UV H$_2$ lines may be hampered by optical depth effects,
that is, by self-absorption in the lines: When the levels are thermally populated, this leads to
higher populations for the lower vibrational levels. Therefore, the lower vibrational
levels have a higher probability of absorbing a fluoresced photon along the line of
sight before the photon can escape from the protoplanetary disk. It might be re-emitted
out of the line of sight, causing increasing flux loss for increasingly lower
levels, which corresponds to lines at increasingly bluer wavelength. This higher
opacity for bluer lines mimics the effect of extinction, as was first found by
\citet{Wood2002}. \citet{McJunkin2016} addressed this interplay between extinction
and self-absorption with two-step radiative transport calculations. In a first step,
the lines blueward of 1450\,\AA\, are used to determine the self-absorption and a
correction for it. In a second step, all lines are then used to determine $A_{V}$.
Since $A_{V}$ is not applied to the first step, however, the effect of self-absorption
might be overestimated. Even in the case of TW~Hya, where no complication by reddening is expected, the results are ambiguous because \citet{McJunkin2016} found a significant opacity in all lines below 1450\,\AA, while
\citet{Herczeg2006} found that progressions [1,4] and [0,1] only have a significant optical depth
for lines below 1300 and 1350\,\AA, respectively. Moreover, \citet{Ardila2002} found that the H$_{2}$ levels for different stars with
$\nu''\ge$4 are optically thin. The latter finding would allow us to use all of
our lines in the analysis, but following \citet{Herczeg2004} would suggest that we omit the bluest line of the [0,1] progression. The study by  \citet{McJunkin2016} would imply
that the two bluest lines of progressions [0,1] and [0,2] are affected
by non-negligible line opacity, and that to a smaller extent, even progressions [1,7] and
[1.4] may be affected because their bluest lines are located around 1450\,\AA.

To gain further insight without applying full radiative transfer calculations, we estimated the optical depth $\tau_{lu}$ in the lines of the four progressions by using Eqs. 1 -- 12 from \citet{McJunkin2016} with the Voigt profile of all lines set to unity. We took the lines in the four progressions into account as well as their pumping transitions near Ly$\alpha$, and we neglected all other H$_{2}$ lines. Considering only the pumping lines (since their level population numbers are expected to be much higher than the population numbers of the other transitions) and under the assumption of $\tau=1$, we calculated the total column density $\log(N_{tot})\sim 15.5$ for an assumed excitation temperature $T_{exc}=2000$\,K. \citet{McJunkin2016} found average values of  $\log(N_{tot})=19.0$ (with DR~Tau as an outlier with $\log(N_{tot})=15.1$) and $T_{exc}=1500$\,K (with TW~Hya as an outlier with $T_{exc}=2500$\,K).
This discrepancy of our high-excitation temperatures leading to rather low column densities can be explained by our omission of all pumping lines blueward of the Ly$\alpha$ line center, which should have even higher population numbers. Moreover, a realistic Voigt profile would lead to higher column densities, as would a $\tau_{lu}$ for the pumping lines higher than one. We therefore consider our $N_{tot}$ value as a lower limit, although it increases (slightly) with temperature.

In a second step, we then computed $\tau_{lu}$ in the individual (fluorescent) lines using $N_{tot}$. We list the $\tau_{lu}$ values we estimated under these assumptions in Table~\ref{tab:lines}. As a result of our simplifications, these numbers should be treated with care, but we are positive that we recover the trend, confirming that progression [0,2] is much more affected than progressions [1,7] and [1,4]. Since the bluest line of progression [0,2] is most affected, this progression indeed cannot be used because other lines cannot be used due to blends.

With these general considerations in mind, we wished to rely only on the measured line flux and refrained from performing full
radiative transport calculations to estimate the line opacity effects. Our motivation was
to infer the lines that may be influenced by self-absorption from our derived
$A_{V}$ and $R_{V}$ values of the different progressions for the individual stars.

\section{Results}\label{sec:results}
From our simulations, we expected to be able to recover the essentially zero extinction of TW~Hya and the other low $A_V$ stars. 
For the other stars, measuring $R_V$ independently from  $A_{V}$ may not lead to significant results  given the noise level of the data and the degeneracy between $R_V$ and $A_{V}$.
Still, assuming $R_{V}$=3.1, we expected to be able to measure $A_{V}$.
We list our results for all stars in Table~\ref{tab:results} and show our results graphically for a representative selection of stars, like in the following example of TW~Hya in the figures published on Zenodo.

\begin{figure*}
\begin{center}
\includegraphics[width=0.85\textwidth, clip]{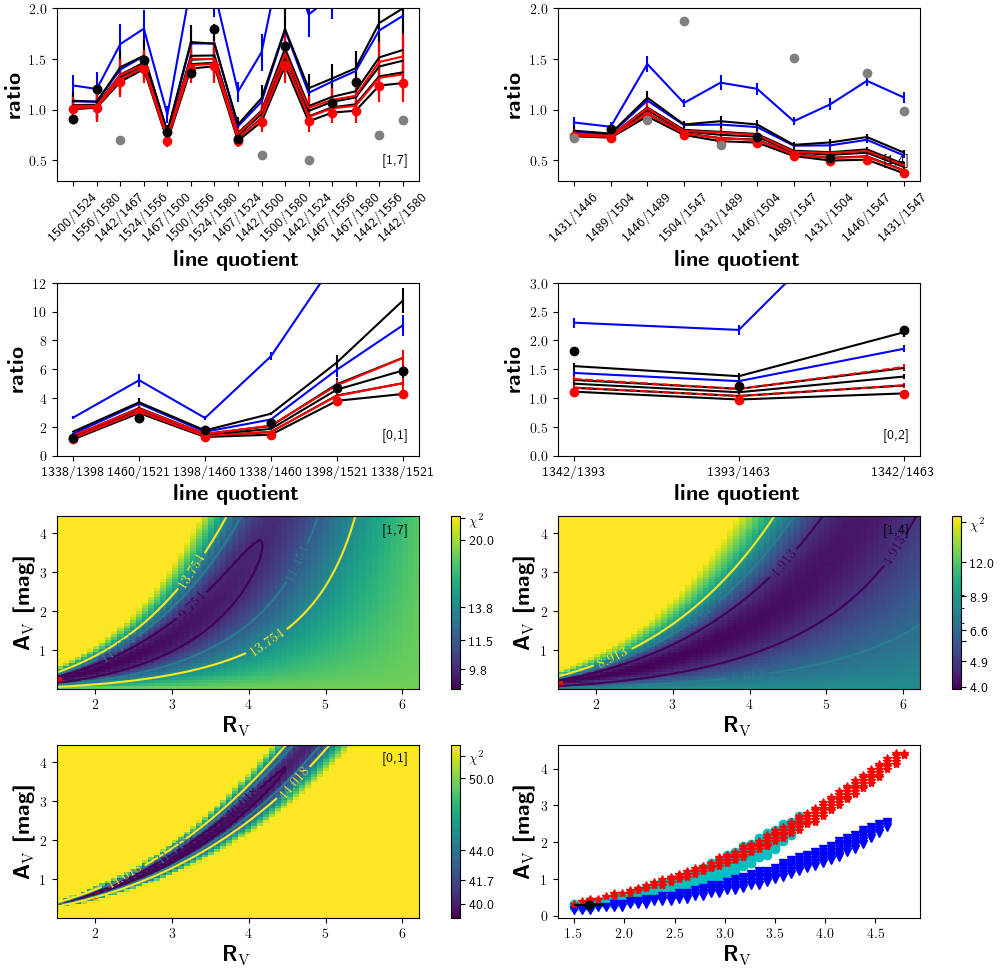}\\%\vspa4ce{-0.4mm}\\
\caption{\label{av_twhya} Determination of  A$_{\mathrm{V}}$ for TW~Hya.
  We show the progressions as indicated in the lower or upper right corner of each panel.
  The top
  four panels show the measured and dereddened flux ratios (red dots)
  and the theoretical
  flux ratios (black dots if the ratio is used for the analysis, gray otherwise).
  Each black line corresponds to the curve defined by
  the dereddend measured ratios according to $R_V$=3.1 and $A_V$=0.0,0.5,1.0,1.5,3.0. The red and blue lines denote the same for $R_V$=2.1 and 4.1 for $A_V$=1.0 and 3.0, respectively
  (the vertical bars are the dereddened errors of the ratios).
  %The black lines indicate the best fit models mentioned in Table~\ref{tab:fit}.
  Three of the four lower panels show
  the corresponding contour plots of the best-fit models in the  $R_{V}$ -- $A_{V}$
  plane. We show the 68\%, 90\%, and 95\% confidence interval as a black, blue, and yellow line,
  respectively. The numbers in these lines state the $C\sim\chi^{2}$s for the respective
  confidence level and are also given in the color bar. The red dot marks the position of the minimum of the distribution.
  In the bottom right panel, we overplot the 180 best-fit $C$s
  for every progression. Progression [1,7] is represented by cyan dots,
  progression [1,4] is represented by blue triangles, and progression [0,1] is represented
  by red stars. The black dot with error bars marks the central position and extent of
  the search box, and it marks the overall best $R_{V}$ -- $A_{V}$ combination.
}
\end{center}
\end{figure*}

We present our results for TW~Hya graphically in Fig.~\ref{av_twhya} as an instructive
example. Similar to Fig.~\ref{testsim1},  we show in the upper four panels the measured and
the theoretical ratios for each progression.  We
realized that all theoretical ratios involving the line at 1442\AA\, from progression
[1,7] are below the measured ratios, and therefore, the measured fluxes must be erroneous, because they cannot be
reached by any dereddening, as shown by the colored curves. Since the line is
near the beginning of an echelle order of the STIS spectrograph, we omitted it from the analysis %(and show the
%theoretical ratios in Fig.~\ref{av_twhya} therefore in gray, while the theoretical
%ratios used for the analysis are shown in black).
The same is true for the line at 1547\AA\, from progression [1,4], which is severely
blended with \ion{C}{iv}. Fig.~\ref{av_twhya} shows for progression [1,7]
that two additional theoretical values are
below the measured values. Both ratios involve the line (1-7)R(6) at
1500.45\AA, which shows a satisfying fit. Since the deviation is only small and within the
3$\sigma$ error for the measured values, we still used the ratios.

%We list in Table\ref{tab:fit} the four to six best values obtained and look for similar
%best fit values in the different progressions, since we know from the simulations that
%the stochastically errors may lead to different distributions of $R_{V}$ and E(B-V) for
%the different progressions, where the overlap mark the correct value. Progression [1,7]
%favours E(B-V)=0.2-0.27
%with $R_{V}$=1.6-2.1, similar to progression [0,1], which favours E(B-V)=0.25-0.33 and
%$R_{V}$=1.6-2.1.
%On the other hand progression [0,2] favours E(B-V)=0.3-0.33 and $R_{V}$=1.1, while
%progression
%[1,4] favours E(B-V)=0.15-0.2 and $R_{V}$=2.1-2.6. For TW~Hya there is a second group of
%similar values with slightly higher $R_{V}$ and $A_{V}$. Since the results from progression
%[0,2] differs significantly from the other progressions, we exclude it from finding an
%overall $A_{V}$ value.

For TW~Hya, we find an overall $A_{V}$=0.3$\pm$0.1\,mag and $R_{V}$=1.65$\pm$0.15.
We therefore conclude that our analysis for TW~Hya favors a low but nonzero
$A_{V}$ in combination with a very steep $R_{V}$. 

For the other six stars with low $A_{V}$
literature values, we find best-fit values of $A_{V}=0.0$\,mag (for a wide variety of formal $R_V$ values for the individual stars). $A_V$ at $R_V$=3.1 agrees with 0.0\,mag within one standard deviation for all these stars (except for TW~Hya).
%values above zero are statistically tolerable, but require  low $R_{V}$ value though:
%$R_{V}$=2.25$\pm$0.5 for DM~Tau, 1.5$\pm$0.3 for CVSO~90, and 1.75$\pm$0.5 for RECX~15. For the results for DM~Tau we did not consider the progression [0,1], since it totally disagreed with the other two progressions.
We show the obtained fits in figures published on Zenodo (see Table~\ref{tab:results}).

The stars with nonzero $A_V$ fall in three categories depending on the number of progressions that can be used to infer $A_V$ (which we therefore list in Table~\ref{tab:results}): In the first category, all three progressions led to about the same result. In the second category, the [0,1] progression led to deviating results and was omitted because the deviation of this progression might be
self-absorption. In the third category, all three progressions led to differing results. In this last category, we found six stars for which the fit did not lead to an overlapping result region, and we were unable to assign an $A_V$ and $R_V$ value to these stars. Nevertheless, we computed a mean $A_V$ value for the case of an assumed $R_V=3.1$. We did this also for all other stars to allow for a better comparison to the literature values.

%For SY~Cha we show our results graphically in Fig.~\ref{av_sycha}.
%We find $A_{V}$=0.6$\pm$0.4\,mag and $R_{V}$=2.1$\pm$0.25.
%For DK~Tau~A the fit for the [0,1] 1521\,\AA\, line is unstable and yields results variable
%by about a factor of two. We therefore omit this line from the A$_{\mathrm{V}}$
%fitting process.
%For the fitting of DK~Tau~A we could not use  for unknown
%reasons also the lines [0,7] 1442\,\AA, [1,7] 1467\,\AA, and [1,4] 1446\,\AA. This leaves us
%for all three progressions only with three usable lines, though the Gaussian line fits
%look acceptable for these lines. We note, that some of the lines of DK~Tau~A exhibit a
%blue wing, among them the line [1,4] 1446\AA. We show the fitting results in
%Fig.~\ref{av_dktau}.
%Nevertheless, progressions [1,7], [1,4], and [0,1] favour a clear overlapping area
%and we find an overall $A_{V}$=1.9$\pm$0.1\,mag and $R_{V}$=3.1$\pm$0.1.
%For MY~Lup we show our results in Fig.~\ref{av_mylup}.
%Again, the fit for the weak [0,1] 1521\,\AA\, line is not usable like for DK~Tau~A.
%We find overall best fit values of $A_{V}$=2.25$\pm$0.25\,mag and $R_{V}$=3.5$\pm$0.25.

%\textbf{Also, for a number for stars, the progression [0,1] led to disagreeing results, because this progression led usually to results located above that of the other progressions in the $R_V$-$A_V$ plane. In that case, we excluded the progression from our box search and indicate the number of used progressions in Table~\ref{tab:results}.}

\begin{table}
\caption{\label{tab:results} Best-fit results for $A_V$ and $R_V$}
\footnotesize
\begin{tabular}[h!]{lcccc}
\hline
\hline
\noalign{\smallskip}
Object & $A_V$ & $R_V$ & no of  &   $A_V$ for \\
       & [mag] &       &progr.& $R_V$=3.1\\
\noalign{\smallskip}
\hline
\noalign{\smallskip}
TW Hya &  0.3$\pm$0.1 & 1.65$\pm$0.15& 3& 1.2$\pm$0.3\\
DM Tau *&  0.0$\pm$0.2 & 2.25$\pm$0.5 & 2& 0.3$\pm$0.3 \\
RECX 11 *& 0.0$\pm$0.2 & 4.0$\pm$0.2 & 2 & 0.0$\pm$0.1\\
RECX 15  *& 0.0$\pm$0.4&1.75$\pm$0.5 & 2&  0.5$\pm$0.5\\
CVSO 90  *& 0.0$\pm$0.4& 1.5$\pm$0.3 & 3&  0.0$\pm$0.2\\
Sz 77 *& 0.0$\pm$0.2 & 5.0$\pm$0.3 & 2 & 0.0$\pm$0.1\\
Sz 111 *& 0.0$\pm$0.3 & 2.25$\pm$0.5 & 2 & 0.2$\pm$0.2\\
\noalign{\smallskip}
\hline
\noalign{\smallskip}
AA Tau  & -- & -- & 0 & 0.7$\pm$0.2\\
CS Cha & 0.6$\pm$0.3 & 3.25$\pm$0.5 & 2 & 0.4$\pm$0.2\\
DE Tau & 3.0$\pm$0.0& 4.25$\pm$0.2 & 2 & 1.0$\pm$0.1\\
DK Tau A *& 1.9$\pm$0.1&3.1$\pm$0.1& 3 & 1.8$\pm$0.2\\
DR Tau & -- & -- & 0 & 1.5$\pm$0.7\\
HN Tau & -- & -- & 0 & 1.1$\pm$0.6\\
MY Lup *&  2.25$\pm$0.25& 3.5$\pm$0.25& 3 & 1.5$\pm$0.2\\
RY Lup & 2.0$\pm$0.5& 4.25$\pm$0.5& 3 & 0.8$\pm$0.2\\
SY Cha *&   0.6$\pm$0.1& 2.75$\pm$0.1 & 2& 0.8$\pm$0.2\\
UX Tau A & 0.6$\pm$0.2&3.25$\pm$0.5 & 3 & 0.5$\pm$0.2\\
XX Cha & 0.2$\pm$0.3 & 2.25$\pm$0.4 & 2 & 0.6$\pm$0.2\\
Sz 45  & 0.4$\pm$0.3 & 3.0$\pm$0.3 & 2 & 0.5$\pm$0.2\\
Sz 66 &  1.0$\pm$0.3 & 3.25$\pm$0.5 & 2 & 0.9$\pm$0.2\\
Sz 68 &  2.4$\pm$0.3&3.25$\pm$0.5& 2 & 1.5$\pm$0.4\\
Sz 71 &  0.6$\pm$0.2& 3.25$\pm$0.2& 2 & 0.5$\pm$0.1\\
Sz 75 & -- & --& 0 & 1.5$\pm$0.5\\
Sz 98 & 1.8$\pm$0.4& 4.0$\pm$0.5& 2 & 0.7$\pm$0.1\\
Sz 102 & 0.2$\pm$0.4 & 2.0$\pm$0.5 & 2 & 0.9$\pm$0.4\\
Sz 103 & 0.2$\pm$0.3 & 2.75$\pm$0.3 & 2 & 0.5$\pm$0.2\\
Sz 114 & -- & -- & 0 & 0.5 $\pm$0.2\\
Sz 129 & -- & -- & 0 & 0.4$\pm$0.1\\
CHX 18n & 0.8$\pm$0.3 & 3.25$\pm$0.3 & 2 & 0.7$\pm$0.1\\
CVSO 146 & 0.2$\pm$0.2 & 2.25$\pm$0.3 & 2 & 0.5$\pm$0.2\\
LkCa 15 & 0.2$\pm$0.2 & 2.75$\pm$0.3 & 2 & 0.4$\pm$0.1\\
RXJ1556& 2.6$\pm$0.3& 4.25$\pm$0.5 & 2 & 0.8$\pm$0.1\\
sstc2dj1608 & 0.0$\pm$0.4 & 2.0$\pm$0.4 & 2 & 0.3$\pm$0.3\\
sstc2dj1612 *& 1.2$\pm$0.4 & 3.5$\pm$0.4 & 2 & 0.9$\pm$0.2\\
\hline
\end{tabular}
\tablebib{Note: For all stars whose names are marked by an asterisks, we show the results graphically in the figures published on Zenodo under \url{https://zenodo.org/records/14034355}.}
\normalsize
\end{table}

\section{Discussion}\label{sec:discussion}

%\subsection{Comparison of the $A_{V}$ and $R_{V}$ measurements to literature}

\begin{figure}
\begin{center}
\includegraphics[width=0.50\textwidth, clip]{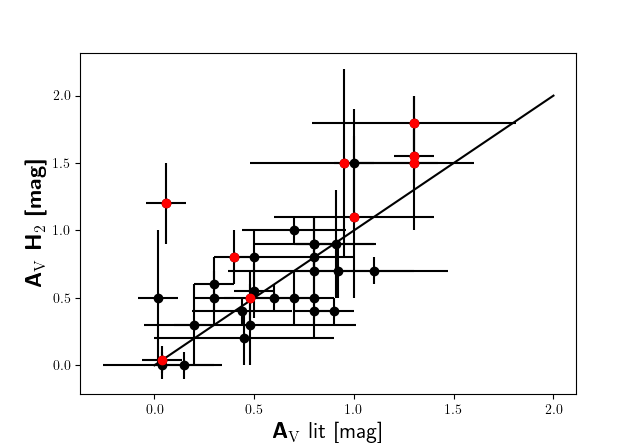}\\%\vspa4ce{-0.4mm}\\
\includegraphics[width=0.50\textwidth, clip]{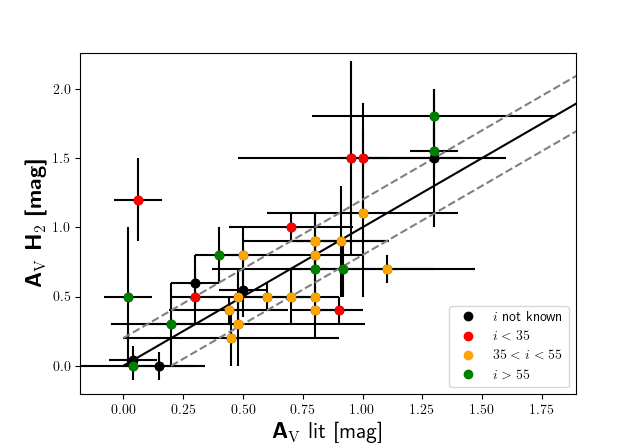}%\vspa4ce{-0.4mm}\\
\caption{\label{comparison} Comparison of our $A_V$ values for fixed $R_V=3.1$ to the literature values. \emph{Top:} When all three progressions could be used to infer the mean $A_V$, the stars are represented by red dots, and when two progressions could be used, they are shown with black dots (progression [0,1] omitted). The black line indicates identity. The four stars with literature values of $A_V$=0.0\,mag are offset by 0.02\,mag for clarity of their error bars. \emph{Bottom:} Same as above, but color-coded with the disk inclination. The dashed gray lines indicate errors of 0.2\,mag.}

\end{center}
\end{figure}

We show our mean $A_V$ values derived for $R_V$=3.1 in a comparison to literature $A_V$ values obtained for the same $R_V$ in Fig.~\ref{comparison}. The Pearson correlation coefficients are $r=0.70$ and $p=3.5\times 10^{-6}$, which indicates a  correlation. The correlation is even better when we omit TW~Hya, which would leave us with $r=0.80$ and $p=2.7 \times 10^{-8}$. TW~Hya is special in the sense that it strongly deviates between the best $A_V$ values assigned by the box fit of $0.3\pm0.1$, with $R_V$=1.65 and the best mean $A_V$=1.2 for $R_V$=3.1. \citet{McJunkin2016} also reported it as one of two stars for which the optical depth affected the lines up to 1450\,\AA. This may also slightly influence our progressions [1,7] and [1,4], which would (incorrectly) lead to higher $A_V$ values. \citet{Herczeg2004} found an $A_V$=1.3\,mag for TW~Hya by ignoring the optical depth effects. This strongly suggests that for TW~Hya, apart from the degeneracy, our results seem to be influenced by self-absorption, which we could not identify for this star because progressions [1,7] and [1.4] may also be influenced.

For the other six stars with literature $A_V=0$\,mag, we also find very low to zero $A_V$ values with the box search method, as for a mean $A_V$ at $R_V$=3.1.  For three stars, low $R_V$ values are preferred by the box search, which may be caused at least partly by the degeneracy of $A_V$ and $R_V$, since we do not find $R_V<2.0$ for the stars with nonzero $A_V$.
Moreover, for five of these seven stars, progression [0,1] leads to significantly higher $A_V$ values. This suggests a contamination by the optical depth for these stars. \citet{McJunkin2016} included TW~Hya, DM~Tau, RECX-11, and RECX-15 in their study and found strong optical depth effects for wavelengths up to 1450\,\AA\,  for these stars. After a correction for optical depth, they inferred high $R_V$ values for these stars, and in particular, for TW~Hya also a high $A_V=2.0$. We agree with \citet{McJunkin2016} for the high optical depth in DM~Tau, RECX-11, and RECX-15, but were unable to find indications of an optical depth in the H$_{2}$ lines for TW~Hya (which may be affected most, which means that our procedure failed; see above). Furthermore, our box search method prefers low $R_V$ values, which is in contrast to what \citet{McJunkin2016} found, except for RECX-11, where we find a comparably high $R_V$ value. 

For the stars with nonzero $A_V$, we have six stars in common with \citet{McJunkin2016}: AA~Tau, DE~Tau, DR~Tau, HN~Tau, LkCa~15, and UX~Tau. For the four stars AA~Tau, DE~Tau, HN~Tau, and LkCa~15, the [0,1] progression leads to higher $A_V$ values, but for AA~Tau and HN~Tau, all three progressions differ. This is also the case for DR~Tau, but progression [0,1] does not lead to the highest $A_V$ values and therefore does not indicate any optical depth effects. For UX~Tau, we find the $A_V$ values of progression [0,1] to be only slightly higher than those of the other two progressions.\citet{McJunkin2016} found that optical depth played a role for five of these stars only for a wavelength below $\sim$1300\,\AA, while for UX~Tau, the affected region extended up to 1350\,\AA. The former would not affect any of our lines, while the latter would affect the bluest line of progression [0,1]. It is also curious that the stars for which \citet{McJunkin2016} found the weakest influence of the optical depth are the stars for which in our analysis all three progressions differ. This discrepancy for the individual stars remains unexplained. Spectra with a higher resolution would 
help us to distinguish between problems introduced by the method or physical 
properties, such as H$_{2}$ emission, which partly originates in different places, for 
example, in the disk wind.

Fig.~\ref{comparison} shows that our $A_V$ values, for which we used three progressions, are all equal to or higher than the literature values, which can be caused by the unattended self-absorption in the [0,1] progression. Of the stars for which we used only two progressions, slightly more than half exhibit a lower $A_V$ than the literature value, and we therefore argue that they are usually not influenced by optical depth.

%\textbf{Although there is rough agreement between the $A_V$ values obtained with the H$_{2}$ lines and literature values, slightly more than half of our $A_V$ values are higher than the literature values found mostly from optical spectra. This may  hint at unattended optical depth effects in our method, or it suggests some additional (UV) extinction in the vicinity of the star. }
%Generally our $A_{V}$ values for all four stars are slightly higher than the ones from
%literature. This may suggest, that in the UV one observes additional extinction in the
%vicinity of the star.

An interesting question is whether H$_{2}$ extinction measurements lead to different results than extinction measurements obtained from the optical for some physical reason. This may be expected because we measure extinction toward the origin of the H$_{2}$ emission in the disk, while the measurements obtained from optical data determine extinction toward the star. We therefore  show in Fig.~\ref{comparison} the extinction values color-coded with disk inclination, where low values indicate face-on disks. We find a tendency for face-on disks to show higher H$_{2}$ measured than in the literature extinction. This might indicate that the light generated in the hot disk gas must pass the outer rims of the disk, while the light from the star is unaffected by this additional extinction. However, we only have three stars (TW~Hya, DK~Tau~A, both with a face-on disk, and DE~Tau) for which our extinction value is 3$\sigma$ higher than the literature value.

\section{Conclusions}\label{sec:conclusion}

We derived $A_{V}$ and $R_{V}$ values for 34 CTTSs  using
measurements of fluoresced H$_{2}$ line fluxes. Using simulated data, we
showed that it is possible to recover $A_V$ and $R_V$ values for the 
typical S/N of the UV data using stars with literature values of $A_{V}$ ranging from 0 to 1.6\,mag. 

%The two different methods differ slightly in their approach of recovering $A_V$: Method I uses the branching ratios and line fluxes from the different progressions to directly obtain $A_{V}$ and $R_{V}$ as well as the flux in each progression while method II 
Our method used the
theoretical line ratios separately for the individual progressions and combined the results from the individual progressions afterward.  
%We find that  Method I with a fixed $R_{V}=3.1$ results in $A_{V}$-values, which are typically higher (especially for TW~Hya) than the
%literature values except for MY~Lup. 
%Similarly, for a fixed $R_V=3.1$,  
%with Method II we also find $A_V$-values at the upper range of the literature values (SY~Cha, DK~Tau, MY~Lup), only TW~Hya shows much higher 
%$A_{V}$ values.
It allowed us to identify lines that are incompatible with the expected ratio pattern and therefore may be prone to systematic flux measurement errors, for example, by a blend with another line. We computed $C$ values (similar to  $\chi^{2}$ values) for the individual progressions
on an $A_{V}$ and $R_{V}$ grid and applied a box search to identify the best-fitting models. This also partly allowed us to break the degeneracy between
$A_{V}$ and $R_{V}$ and to obtain results for the $A_V$ and $R_V$ parameters. 

In this way, we found low $A_V$ and $R_V$ values for the stars with $A_V$=0.0\,mag in the literature. For TW~Hya and fixing $R_V$=3.1, we found a significant $A_V$=1.2$\pm0.3$\,mag, without any indications of contamination by self-absorption. Although this may indicate that for TW~Hya all our progressions are affected by optical depth effects, it  may also indicate additional UV extinction in the vicinity of the star because for most other stars with face-on disks our $A_V$ values are also higher than the literature values.

%For TW~Hya, we obtain overall best fit values of $A_{V}$=0.3\,mag and
%$R_{V}$=1.65, which is higher and steeper compared to literature values,
%which are usually $A_{V}$=0\,mag. This may be caused by the FUV photons being more sensitive to smaller
%grain sizes than longer wavelength, but it may also be a problem of self-absorption in the
%H$_{2}$ lines at bluer wavelength. 
%For TW~Hya different calculations exist, at which wavelength
%the opacity should not be negligible any more, but they differ significantly, favouring
%1450\AA\, \citep{McJunkin2016} or only
%1300\AA\, \citep{Herczeg2004} as threshold. In the latter case, no lines we use would be
%affected and our result would be significant, while in the first case  lines
%of the progressions  [0,1] and [0,2] would be affected. Progression [0,2] has only two usable lines for many of our stars and it leads to deviating results, which may also be caused by the lower variance of the theoretically expected flux ratios, which should make the progression the least sensitive one. On the other hand progression [0,1] should be the most sensitive one and therefore should also react strongly to self-absorption, which is apparently the case for some stars, where it leads to a steeper $A_V$--$R_V$ distribution. There we found systematic deviations of the $C$ distribution for progression [0,1] compared to progressions [1,7] and [1.4].  Therefore, the influence of self-absorption stays a little bit elusive, but should not play a significant role here, since we omit progression [0,1], where it leads to lower $R_V$ and higher $A_V$ values.

For stars with nonzero $A_V$ literature values, we found lower as well as higher $R_V$ values than the canonical $R_V$=3.1. For a mean $A_V$ obtained from two or three progressions at $R_V$=3.1, we found reasonable agreement with literature values. 

For the progressions we used, we found that progression [0,2] cannot be used due to a combination of unusable lines and because it is affected by self-absorption. For many stars, the same is true for progression [0,1], which was excluded from the $A_V$ determination when it exhibited systematically steeper $R_V$-$A_V$ distributions than the other progressions. Progressions [1,4] and [1,7] seem to be usable for all stars (except perhaps for TW~Hya).

In summary, the exact $A_V$ values that we derived may be affected by some systematic uncertainties such as optical depth effects. However, the values that result from the simple reconstruction are quite close to the literature values and only allow some but not much additional UV extinction in the vicinity of the warm disk, where the H$_{2}$ fluorescent lines are created. 

We therefore argue that the H$_2$-derived $A_V$ values, and the literature $A_V$ for that matter, are good estimates of the extinction toward the inner warm disk. The general agreement between the literature and our H$_2$-based $A_V$ values suggests that no (significant) additional extinction sources exist between the star and the disk. 

We found no indication for additional UV extinction either, for instance, by steep extinction laws. 
This is particularly relevant when modeling the UV spectra of accreting stars because our values provide an estimate of the UV-$A_V$ that cannot be exceeded. Specifically in the context of models that describe the stellar UV spectrum as a combination of accretion columns, it is possible that some of these accretion fluxes can be immediately absorbed in the vicinity of the star, but our results indicate no additional UV-extinction. Therefore, the H$_2$-based or standard $A_V$ values should also be used to redden the calculated UV emission from the accretion columns for comparison with observations. 

%Therefore,  the H$_2$-based or literature  $A_V$s together with standard extinction laws can be used to deredden the stellar fluxes.  

In the future, better (higher S/N) UV H$_2$ line fluxes together with simultaneously treated extinction and optical depth will eventually allow one to derive the extinction toward the inner disk.

\section{Data availability}\label{data-avail}
The figures published via Zenodo can be downloaded under \url{https://zenodo.org/records/14034355}

\begin{acknowledgements}
  The authors acknowledge funding
 through Deutsche Forschungsgemeinschaft DFG program IDs EI~409/20-1
and SCHN~1382/4-1 in the framework of the YTTHACA project (Young stars at
T\"ubingen, Tautenburg, Hamburg \& ESO -- A Coordinated Analysis).
Funded by the European Union under the European Union’s Horizon Europe
Research \& Innovation Programme 101039452 (WANDA). Views and opinions
expressed are, however, those of the author(s) only and do not necessarily
reflect those of the European Union or the European Research Council.
Neither the European Union nor the granting authority can be held
responsible for them.
 Based on observations obtained with the NASA/ESA Hubble Space Telescope, retrieved from the Mikulski Archive for Space Telescopes (MAST) at the Space Telescope Science Institute (STScI). STScI is operated by the Association of Universities for Research in Astronomy, Inc. under NASA contract NAS 5-26555.  %  We thank our referee for the careful reading and the suggestions for improvement.
This work
benefited from discussions with the ODYSSEUS team (HST AR-16129, \citet{Espaillat2022}, https://sites.bu.edu/odysseus/) and especially with G.~J.~Herczeg and K. Grankin.   
This work made use of PyAstronomy \citep{Czesla2019pya}, which can be downloaded at {\tt https://github.com/sczesla/PyAstronomy}.

\end{acknowledgements}

\bibliographystyle{aa}
\bibliography{papers}

\begin{thebibliography}{42}
\expandafter\ifx\csname natexlab\endcsname\relax\def\natexlab#1{#1}\fi

\bibitem[{{Abgrall} {et~al.}(1993){Abgrall}, {Roueff}, {Launay}, {Roncin}, \&
  {Subtil}}]{Abgrall1993}
{Abgrall}, H., {Roueff}, E., {Launay}, F., {Roncin}, J.~Y., \& {Subtil}, J.~L.
  1993, \aaps, 101, 273

\bibitem[{{Alcal{\'a}} {et~al.}(2017){Alcal{\'a}}, {Manara}, {Natta}, {Frasca},
  {Testi}, {Nisini}, {Stelzer}, {Williams}, {Antoniucci}, {Biazzo}, {Covino},
  {Esposito}, {Getman}, \& {Rigliaco}}]{Alcala2017}
{Alcal{\'a}}, J.~M., {Manara}, C.~F., {Natta}, A., {et~al.} 2017, \aap, 600,
  A20

\bibitem[{{Ardila} {et~al.}(2002){Ardila}, {Basri}, {Walter}, {Valenti}, \&
  {Johns-Krull}}]{Ardila2002}
{Ardila}, D.~R., {Basri}, G., {Walter}, F.~M., {Valenti}, J.~A., \&
  {Johns-Krull}, C.~M. 2002, \apj, 566, 1100

\bibitem[{{Calvet} {et~al.}(2005){Calvet}, {Brice{\~n}o}, {Hern{\'a}ndez},
  {Hoyer}, {Hartmann}, {Sicilia-Aguilar}, {Megeath}, \&
  {D'Alessio}}]{Calvet2005}
{Calvet}, N., {Brice{\~n}o}, C., {Hern{\'a}ndez}, J., {et~al.} 2005, \aj, 129,
  935

\bibitem[{{Cardelli} {et~al.}(1989){Cardelli}, {Clayton}, \&
  {Mathis}}]{Cardelli1989}
{Cardelli}, J.~A., {Clayton}, G.~C., \& {Mathis}, J.~S. 1989, \apj, 345, 245

\bibitem[{{Carvalho} \& {Hillenbrand}(2022)}]{Carvalho2022}
{Carvalho}, A.~S. \& {Hillenbrand}, L.~A. 2022, \apj, 940, 156

\bibitem[{{Czesla} {et~al.}(2019){Czesla}, {Schr{\"o}ter}, {Schneider},
  {Huber}, {Pfeifer}, {Andreasen}, \& {Zechmeister}}]{Czesla2019pya}
{Czesla}, S., {Schr{\"o}ter}, S., {Schneider}, C.~P., {et~al.} 2019, {PyA:
  Python astronomy-related packages}

\bibitem[{{Draine}(2003)}]{Draine2003}
{Draine}, B.~T. 2003, \araa, 41, 241

\bibitem[{{Espaillat} {et~al.}(2022){Espaillat}, {Herczeg}, {Thanathibodee},
  {Pittman}, {Calvet}, {Arulanantham}, {France}, {Serna}, {Hern{\'a}ndez},
  {K{\'o}sp{\'a}l}, {Walter}, {Frasca}, {Fischer}, {Johns-Krull}, {Schneider},
  {Robinson}, {Edwards}, {{\'A}brah{\'a}m}, {Fang}, {Erkal}, {Manara},
  {Alcal{\'a}}, {Alecian}, {Alexander}, {Alonso-Santiago}, {Antoniucci},
  {Ardila}, {Banzatti}, {Benisty}, {Bergin}, {Biazzo}, {Brice{\~n}o},
  {Campbell-White}, {Cleeves}, {Coffey}, {Eisl{\"o}ffel}, {Facchini}, {Fedele},
  {Fiorellino}, {Froebrich}, {Gangi}, {Giannini}, {Grankin}, {G{\"u}nther},
  {Guo}, {Hartmann}, {Hillenbrand}, {Hinton}, {Kastner}, {Koen}, {Mauc{\'o}},
  {Mendigut{\'\i}a}, {Nisini}, {Panwar}, {Principe}, {Robberto},
  {Sicilia-Aguilar}, {Valenti}, {Wendeborn}, {Williams}, {Xu}, \&
  {Yadav}}]{Espaillat2022}
{Espaillat}, C.~C., {Herczeg}, G.~J., {Thanathibodee}, T., {et~al.} 2022, \aj,
  163, 114

\bibitem[{{Fitzpatrick}(1999)}]{Fitzpatrick1999}
{Fitzpatrick}, E.~L. 1999, \pasp, 111, 63

\bibitem[{{France} {et~al.}(2023){France}, {Arulanantham}, {Maloney}, {Cauley},
  {{\'A}brah{\'a}m}, {Alcal{\'a}}, {Campbell-White}, {Fiorellino}, {Herczeg},
  {Nisini}, \& {Vioque}}]{France2023}
{France}, K., {Arulanantham}, N., {Maloney}, E., {et~al.} 2023, \aj, 166, 67

\bibitem[{{Gangi} {et~al.}(2023){Gangi}, {Nisini}, {Manara}, {France},
  {Antoniucci}, {Biazzo}, {Giannini}, {Herczeg}, {Alcal{\'a}}, {Frasca},
  {Mauc{\'o}}, {Campbell-White}, {Siwak}, {Venuti}, {Schneider},
  {K{\'o}sp{\'a}l}, {Caratti o Garatti}, {Fiorellino}, {Rigliaco}, \&
  {Yadav}}]{Gangi2023}
{Gangi}, M., {Nisini}, B., {Manara}, C.~F., {et~al.} 2023, \aap, 675, A153

\bibitem[{{Grankin}(2017)}]{Grankin2017}
{Grankin}, K.~N. 2017, in Astronomical Society of the Pacific Conference
  Series, Vol. 511, Non-Stable Universe: Energetic Resources, Activity
  Phenomena, and Evolutionary Processes, ed. A.~M. {Mickaelian}, H.~A.
  {Harutyunian}, \& E.~H. {Nikoghosyan}, 37

\bibitem[{{Green} {et~al.}(2012){Green}, {Froning}, {Osterman}, {Ebbets},
  {Heap}, {Leitherer}, {Linsky}, {Savage}, {Sembach}, {Shull}, {Siegmund},
  {Snow}, {Spencer}, {Stern}, {Stocke}, {Welsh}, {B{\'e}land}, {Burgh},
  {Danforth}, {France}, {Keeney}, {McPhate}, {Penton}, {Andrews},
  {Brownsberger}, {Morse}, \& {Wilkinson}}]{Green2012}
{Green}, J.~C., {Froning}, C.~S., {Osterman}, S., {et~al.} 2012, \apj, 744, 60

\bibitem[{{Hartigan} \& {Morse}(2007)}]{Hartigan2007}
{Hartigan}, P. \& {Morse}, J. 2007, \apj, 660, 426

\bibitem[{{Herczeg} \& {Hillenbrand}(2014)}]{Herczeg2014}
{Herczeg}, G.~J. \& {Hillenbrand}, L.~A. 2014, \apj, 786, 97

\bibitem[{{Herczeg} {et~al.}(2002){Herczeg}, {Linsky}, {Valenti},
  {Johns-Krull}, \& {Wood}}]{Herczeg2002}
{Herczeg}, G.~J., {Linsky}, J.~L., {Valenti}, J.~A., {Johns-Krull}, C.~M., \&
  {Wood}, B.~E. 2002, \apj, 572, 310

\bibitem[{{Herczeg} {et~al.}(2006){Herczeg}, {Linsky}, {Walter}, {Gahm}, \&
  {Johns-Krull}}]{Herczeg2006}
{Herczeg}, G.~J., {Linsky}, J.~L., {Walter}, F.~M., {Gahm}, G.~F., \&
  {Johns-Krull}, C.~M. 2006, \apjs, 165, 256

\bibitem[{{Herczeg} {et~al.}(2004){Herczeg}, {Wood}, {Linsky}, {Valenti}, \&
  {Johns-Krull}}]{Herczeg2004}
{Herczeg}, G.~J., {Wood}, B.~E., {Linsky}, J.~L., {Valenti}, J.~A., \&
  {Johns-Krull}, C.~M. 2004, \apj, 607, 369

\bibitem[{{Hoadley} {et~al.}(2015){Hoadley}, {France}, {Alexander}, {McJunkin},
  \& {Schneider}}]{Hoadley2015}
{Hoadley}, K., {France}, K., {Alexander}, R.~D., {McJunkin}, M., \&
  {Schneider}, P.~C. 2015, \apj, 812, 41

\bibitem[{{Ingleby} {et~al.}(2013){Ingleby}, {Calvet}, {Herczeg}, {Blaty},
  {Walter}, {Ardila}, {Alexander}, {Edwards}, {Espaillat}, {Gregory},
  {Hillenbrand}, \& {Brown}}]{Ingleby2013}
{Ingleby}, L., {Calvet}, N., {Herczeg}, G., {et~al.} 2013, \apj, 767, 112

\bibitem[{{Jordan} {et~al.}(1977){Jordan}, {Brueckner}, {Bartoe}, {Sandlin}, \&
  {van Hoosier}}]{Jordan1977}
{Jordan}, C., {Brueckner}, G.~E., {Bartoe}, J. D.~F., {Sandlin}, G.~D., \& {van
  Hoosier}, M.~E. 1977, \nat, 270, 326

\bibitem[{{Kenyon} \& {Hartmann}(1995)}]{Kenyon1995}
{Kenyon}, S.~J. \& {Hartmann}, L. 1995, \apjs, 101, 117

\bibitem[{{Luhman} {et~al.}(2017){Luhman}, {Mamajek}, {Shukla}, \&
  {Loutrel}}]{Luhman2017}
{Luhman}, K.~L., {Mamajek}, E.~E., {Shukla}, S.~J., \& {Loutrel}, N.~P. 2017,
  \aj, 153, 46

\bibitem[{{Manara} {et~al.}(2017){Manara}, {Testi}, {Herczeg}, {Pascucci},
  {Alcal{\'a}}, {Natta}, {Antoniucci}, {Fedele}, {Mulders}, {Henning},
  {Mohanty}, {Prusti}, \& {Rigliaco}}]{Manara2017}
{Manara}, C.~F., {Testi}, L., {Herczeg}, G.~J., {et~al.} 2017, \aap, 604, A127

\bibitem[{{Mathis}(1990)}]{Mathis1990}
{Mathis}, J.~S. 1990, \araa, 28, 37

\bibitem[{{McJunkin} {et~al.}(2016){McJunkin}, {France}, {Schindhelm},
  {Herczeg}, {Schneider}, \& {Brown}}]{McJunkin2016}
{McJunkin}, M., {France}, K., {Schindhelm}, R., {et~al.} 2016, \apj, 828, 69

\bibitem[{{McJunkin} {et~al.}(2014){McJunkin}, {France}, {Schneider},
  {Herczeg}, {Brown}, {Hillenbrand}, {Schindhelm}, \& {Edwards}}]{McJunkin2014}
{McJunkin}, M., {France}, K., {Schneider}, P.~C., {et~al.} 2014, \apj, 780, 150

\bibitem[{{Miller}(1968)}]{Miller1968}
{Miller}, J.~S. 1968, \apjl, 154, L57

\bibitem[{{Predehl} \& {Schmitt}(1995)}]{Predehl1995}
{Predehl}, P. \& {Schmitt}, J.~H.~M.~M. 1995, \aap, 293, 889

\bibitem[{{Rieke} \& {Lebofsky}(1985)}]{Rieke1985}
{Rieke}, G.~H. \& {Lebofsky}, M.~J. 1985, \apj, 288, 618

\bibitem[{{Roman-Duval} {et~al.}(2020){Roman-Duval}, {Proffitt}, {Taylor},
  {Monroe}, {Fischer}, {Fischer}, {Fullerton}, {Aloisi}, {Britt}, {Busko},
  {Carlberg}, {De Rosa}, {Jedrzejewski}, {Lockwood}, {Frazer}, {Hernandez},
  {James}, {Oliveira}, {Plesha}, {Riedel}, {Riley}, {Sahnow}, {Sankrit},
  {Shaw}, {Smith}, {Sohn}, {Som}, {Ubeda}, \& {Welty}}]{Ullyses2020}
{Roman-Duval}, J., {Proffitt}, C.~R., {Taylor}, J.~M., {et~al.} 2020, Research
  Notes of the American Astronomical Society, 4, 205

\bibitem[{{Rucinski} \& {Krautter}(1983)}]{Rucinski1983}
{Rucinski}, S.~M. \& {Krautter}, J. 1983, \aap, 121, 217

\bibitem[{{Rugel} {et~al.}(2018){Rugel}, {Fedele}, \& {Herczeg}}]{Rugel2018}
{Rugel}, M., {Fedele}, D., \& {Herczeg}, G. 2018, \aap, 609, A70

\bibitem[{{Schindhelm} {et~al.}(2012){Schindhelm}, {France}, {Herczeg},
  {Bergin}, {Yang}, {Brown}, {Brown}, {Linsky}, \& {Valenti}}]{Schindhelm2012}
{Schindhelm}, R., {France}, K., {Herczeg}, G.~J., {et~al.} 2012, \apjl, 756,
  L23

\bibitem[{{Schneider} {et~al.}(2015){Schneider}, {France}, {G{\"u}nther},
  {Herczeg}, {Robrade}, {Bouvier}, {McJunkin}, \& {Schmitt}}]{Schneider2015}
{Schneider}, P.~C., {France}, K., {G{\"u}nther}, H.~M., {et~al.} 2015, \aap,
  584, A51

\bibitem[{{Schneider} {et~al.}(2018){Schneider}, {Manara}, {Facchini},
  {G{\"u}nther}, {Herczeg}, {Fedele}, \& {Teixeira}}]{Schneider2018}
{Schneider}, P.~C., {Manara}, C.~F., {Facchini}, S., {et~al.} 2018, \aap, 614,
  A108

\bibitem[{{Schultz} \& {Wiemer}(1975)}]{Schultz1975}
{Schultz}, G.~V. \& {Wiemer}, W. 1975, \aap, 43, 133

\bibitem[{{Shull}(1978)}]{Shull1978}
{Shull}, J.~M. 1978, \apj, 224, 841

\bibitem[{{Sternberg}(1988)}]{Sternberg1988}
{Sternberg}, A. 1988, \apj, 332, 400

\bibitem[{{Welty} \& {Fowler}(1992)}]{Welty1992}
{Welty}, D.~E. \& {Fowler}, J.~R. 1992, \apj, 393, 193

\bibitem[{{Wood} {et~al.}(2002){Wood}, {Karovska}, \& {Raymond}}]{Wood2002}
{Wood}, B.~E., {Karovska}, M., \& {Raymond}, J.~C. 2002, \apj, 575, 1057

\end{thebibliography}

\appendix

\section{Total flux in a single progression (Method II)}\label{method1}

Here we discuss an alternative method that is conceptually even simpler than the method introduced above but lacks certain diagnostics, which makes it inferior in the end. This alternative  method is based on the fact that  photon flux in an
individual line, $F^{*}_{ij}$,  is given by the total flux $F[i,k]$ in progression [i,k] by

\begin{equation} F^{*}_{ij} = F[i,k] \cdot B_{ij}, \end{equation}
where $B_{ij}$ is the accompanying branching ratio, defined as
$B_{ij}=A_{ij}/\sum A_{[i,k]}$ with the spontaneous emission Einstein coefficient $A_{ij}$. Since we measure the reddened energy flux $F^{'}_{ij}$ this
leads to 
\begin{equation}    
F^{'}_{ij}=\tau(\lambda;A_{V},R_{V})\cdot F[i,k] \cdot \frac{hc}{\lambda_{ij}}\cdot B_{ij},
\end{equation}
where $\tau(\lambda;A_{V}, R_{V})$
is the transmission for wavelength $\lambda$ due to reddening and $\frac{hc}{\lambda_{ij}}$ converts photon flux to energy flux. For a more detailed discussion we refer to Appendix~\ref{app:method1}. 

For any progression [i,k] and $n$ measured 
line fluxes in that progression, this leads to a set of $n$ equations with the three free parameters:
$F[i,k]$, $R_{V}$ and $A_{V}$; the latter two determine $\tau(\lambda,A_{V},R_{V})$. Therefore, $n\geq3$ lines would be already sufficient to provide an extinction estimate, and more lines can be used to optimize the result using, e.\,g.
the  $\chi^{2}$ method. In fact, only three or four H$_2$ lines provide reliable line fluxes in some progressions. 

The extension from using only one progression to using more progressions is straight forward, since each additional progression 
introduces only one additional unknown to the set of equations as $A_{V}$ and $R_{V}$ 
stay the same. Therefore a large number of progressions with each having
several measured line fluxes is desirable. 

We chose progressions
[1,7], [1,4], and [0,1] for this method, since test calculations for our stars show that progression [0,2] only enlarged
$\chi^{2}$ without altering the results notably. 
Moreover, we use an $A_{V}$ and $R_{V}$ grid for the optimization and fit only the F[a,b] values for each  $A_{V}$ and $R_{V}$-pair;
we then record the $\chi^2$-value for each $A_{V}$ and $R_{V}$-pair and determine the best extinction values together with its confidence range from these $\chi^2$-values.

\begin{figure}
\begin{center}
  \includegraphics[width=0.4\textwidth, clip]{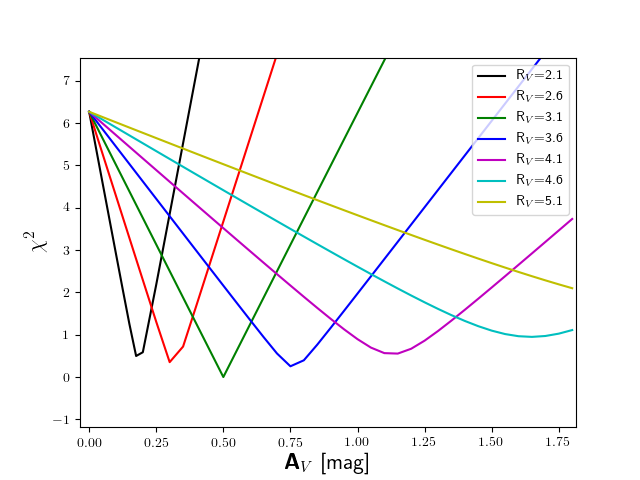}\\%\vspa4ce{-0.4mm}\\
  \includegraphics[width=0.4\textwidth, clip]{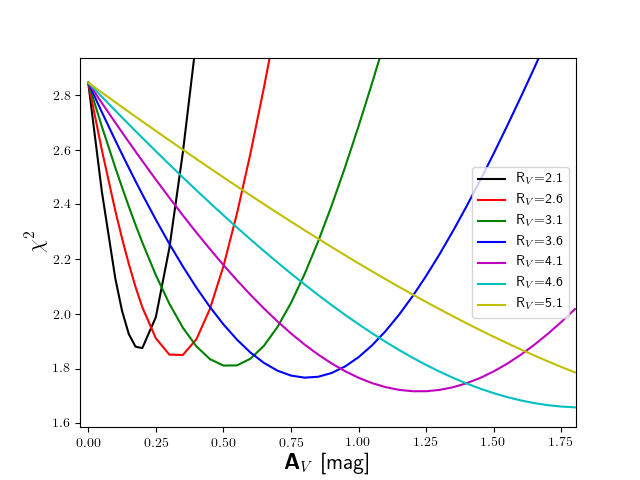}\\%\vspa4ce{-0.4mm}\\
  \includegraphics[width=0.4\textwidth, clip]{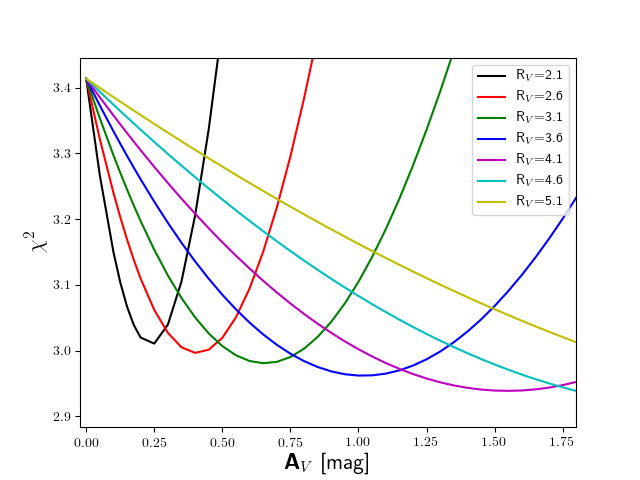}\\%\vspa4ce{-0.4mm}\\
  \includegraphics[width=0.4\textwidth, clip]{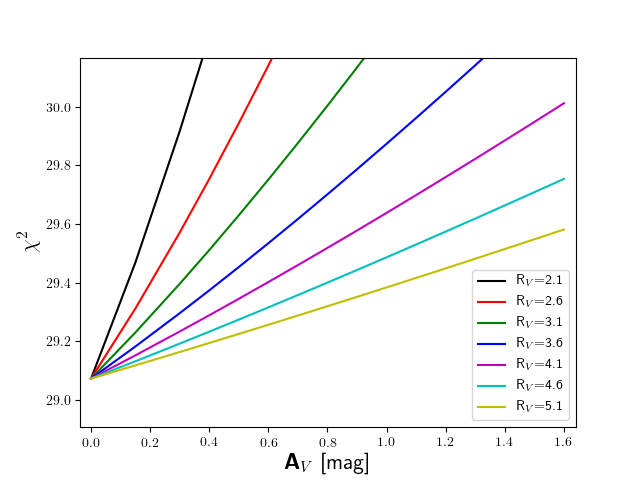}%
  \caption{\label{testsim} Retrieval of $A_{V}$ by $\chi^{2}$ for different $R_{V}$ as
   marked in the legend for simulated data. Input data is $R_{V}$=3.1 and
   $A_{V}$=0.5\,mag except for the lower right panel, where $A_{V}$=0.0\,mag
   \emph{Top:} Relative error of
   0.01, but no statistical noise in the data.
   \emph{Upper middle:} Relative error of 0.05, also noise in the data.
   \emph{Lower middle:} Relative error of 0.10, also noise in the data.
   \emph{Bottom:} Relative error of 0.05, also noise in the data.
}
\end{center}
\end{figure}

To test this approach, we simulated data like for Method I and show the results  in Fig.~\ref{testsim} for random realizations of the `observed' flux vector. Figure~\ref{testsim} shows that small relative errors usually allow us to recover $A_{V}$, even $R_{V}$ for small or negligible measurement errors.
For the data at hand, a relative error of 0.05 best describes the measurements and allows us to recover $A_{V}$ values from 0.0\,mag  to 0.5\,mag, while larger errors introduce subsequently larger uncertainties.

\subsection{Detailed derivation of Method II}\label{app:method1}
 
% \subsection{Equations for line emissivities}
 
The line emissivity (units: erg\,cm$^{-3}$\,s$^{-1}$) of a transition from an upper level $u = \nu',J'$ to an lower level $l=\nu'',J''$ is given by
\begin{equation}\label{EQ_1}
j_{ul} = \frac{1}{4\pi}  n_u\, A_{ul}\, E_{ul}\,,
\end{equation} 
with $A_{ul}$ being the Einstein coefficient, $n_u$ the number density of the energy level $E_u$, and $E_{ul}$ being the energy difference between the two states.
The energy difference for each transition is associated with the release of a photon at wavelength $\lambda_{ul}$ via
\begin{equation}\label{EQ_11}
E_{ul} = \frac{hc}{\lambda_{ul}}
.\end{equation}
At this point we introduce the branching ratio for a progression:
\begin{equation}\label{EQ_2}
B_{ul} = \frac{A_{ul}}{\sum_{[\nu',J']} A_{ul}}\,
,\end{equation} 
where the sum goes over all transitions in the  progression [$\nu',J'$].
$B_{ul}$ can be interpreted as probability of the transition $u\longrightarrow l$, since
\begin{equation}
\sum_{[\nu',J']} B_{ul} = 1\,.
\end{equation}
For $N$ considered transitions, i.e. number of lines in a progression, we can calculate the following expression using Eqs.\,\ref{EQ_1} and \ref{EQ_2}:\begin{equation}\label{EQ_3}
\frac{j_{ul}}{B_{ul}\,E_{ul}} = \frac{1}{4\pi} n_u \sum_{[\nu',J']} A_{ul}
.\end{equation}
Equation\,\ref{EQ_3} gives the same constant for every transition, i.e. it is independent of the considered transition $u\longrightarrow l$
We can add all these expressions for the $N$ transitions to arrive at\begin{equation}
\frac{1}{N}\sum_{[\nu',J']} \frac{j_{ul}}{B_{ul}\,E_{ul}} = \frac{1}{4\pi} n_u\,\sum_{[\nu',J']} A_{ul} .
\end{equation}
By summing up all emissivities divided by their individual branching ratios and individual photon energy we get a constant that depends only on the sum of all relevant Einstein coefficients.

The Method I described here utilises the $N$ equations (Eq. \ref{EQ_3}) in the following way:
The extinction law with parameters $A_V$ and $R_V$ is a wavelength dependent function, here introduced as $\tau(\lambda; A_V, R_V)$, which reduces effectively the intrinsic line emissivities $j_{ul}^{\text{int}}$ to the observed line emissivities $j_{ul}^{\text{obs}}$ via
\begin{equation}
j_{ul}^{\text{obs}} =\tau(\lambda_{ul}; A_V, R_V)\,\cdot\, j_{ul}^{\text{int}} \,.
\end{equation}
Introducing the extinction into Eq.~\ref{EQ_3} leads to a new set of $N$ equations, which are now
\begin{equation}\label{EQ_4}
\frac{1}{4\pi} n_u \sum_{[\nu',J']} A_{ul} = \frac{ j_{ul}^{\text{int}}}{B_{ul}\,E_{ul}} = \frac{ \tau^{-1}(\lambda_{ul}; A_V, R_V)\,\cdot\, j_{ul}^{\text{obs}} }{B_{ul}\,E_{ul}}
.\end{equation}
Incorporating Eq.\,\ref{EQ_11} into Eq.\,\ref{EQ_4} leads to
\begin{equation}\label{EQ_5}
   \frac{ \tau^{-1}(\lambda_{ul}; A_V, R_V)\,\cdot\, j_{ul}^{\text{obs}}\,\cdot \lambda_{ul} }{B_{ul} } =    1/4\pi hc\,\cdot\, n_u \sum_{[\nu',J']} A_{ul} = \text{const}
.\end{equation}
From observations we can measure the $N$ line fluxes and therefore line emissivities (see the next section) and get  a set of $N$ equations with three parameters $A_V, R_V,$ and the constant.

In the following we give an example with 4 measured lines,
where $j_{51}, j_{52}, j_{53}, j_{54}$ are the measured line emissivities, $A_{51}, A_{52}, A_{53}, A_{54}$ are the Einstein coefficients, and $\lambda_{51}, \lambda_{52}, \lambda_{53}, \lambda_{54}$ are the photon wavelengths.
The set of $N=4$ equations are
\begin{align*}
 \frac{ \tau^{-1}(\lambda_{51}; A_V, R_V)\,\cdot\, j_{51}^{\text{obs}}\,\cdot \lambda_{51} }{B_{51} } & = C \\
  \frac{ \tau^{-1}(\lambda_{52}; A_V, R_V)\,\cdot\, j_{52}^{\text{obs}}\,\cdot \lambda_{52} }{B_{52} } & = C \\
   \frac{ \tau^{-1}(\lambda_{53}; A_V, R_V)\,\cdot\, j_{53}^{\text{obs}}\,\cdot \lambda_{53} }{B_{53} } & = C\\
    \frac{ \tau^{-1}(\lambda_{54}; A_V, R_V)\,\cdot\, j_{54}^{\text{obs}}\,\cdot \lambda_{54} }{B_{54} } & = C.
\end{align*}

  If one observes for an individual star a second progression, the constant $C$ changes, but $A_{V}$ and $R_{V}$ stay the same, so that just one additional variable is added and another $N'$ equations, depending on the number $N'$ of observed lines in the second progression.

%  \subsection{Equations for line fluxes}

 In Method II we use line fluxes and not emissivities. Here we show, how one gets
 from the line emissivities $j_{ul}$ (units: erg\,cm$^{-3}$\,s$^{-1}$)  to the integrated line fluxes $F_{ul}$ (units: erg\,cm$^{-2}$\,s$^{-1}$):
The emitting gas is confined to a volume $V_{\text{emit}}$ located at the distance $D$ to the observer. The total energy per unit time in an emission line $u\longrightarrow l$ radiated from the emitting gas is then
\begin{equation}
L_{ul} = \int_{V_{\text{emit}}}\,j_{ul}\,\text{d}V .
\end{equation}
This is the line luminosity (units: erg\,s$^{-1}$). The integrated line flux is then given by
\begin{equation}
F_{ul} = \frac{L_{ul}}{4\pi D^2} = \frac{\int_{V_{\text{emit}}}\,j_{ul}\,\text{d}V}{4\pi D^2}
.\end{equation}
If all emission lines emerge from the same volume at the same distance we have
\begin{equation}
F_{ul} =   \frac{\int_{V_{\text{emit}}} \text{d}V}{4\pi D^2}\,\cdot \,j_{ul} \quad \longrightarrow \quad F_{ul} = \text{const} \,\cdot \,j_{ul}
.\end{equation}
Therefore, we can substitute the line emissivities with the line fluxes,\begin{equation}\label{EQ_6}
   \frac{ \tau^{-1}(\lambda_{ul}; A_V, R_V)\,\cdot\, F_{ul}^{\text{obs}}\,\cdot \lambda_{ul} }{B_{ul} } =    \text{const}
,\end{equation}
holds, though with a different constant.

\subsection{$A_{V}$  derived with Method II}\label{sec:resultsi}
\begin{figure}
\begin{center}
  \includegraphics[width=0.4\textwidth, clip]{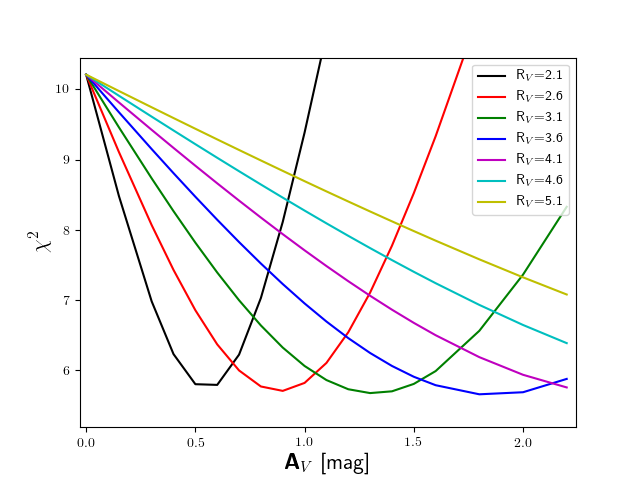}\\%\vspa4ce{-0.4mm}\\
  \includegraphics[width=0.4\textwidth, clip]{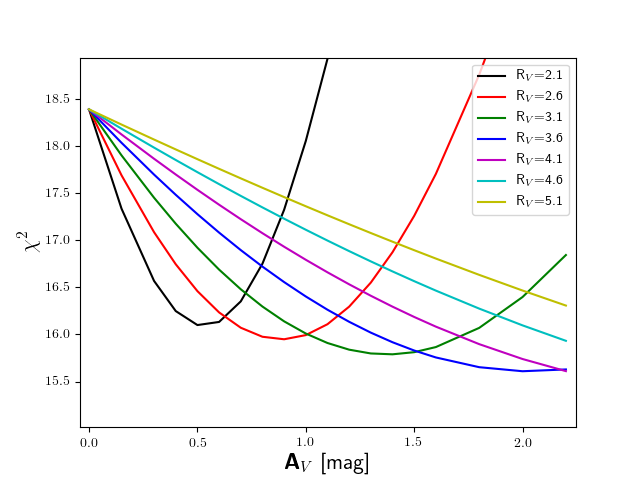}\\%\vspa4ce{-0.4mm}\\
  \includegraphics[width=0.4\textwidth, clip]{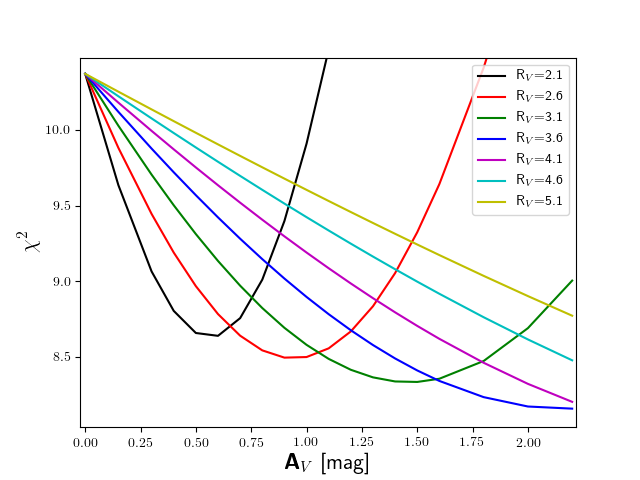}\\%\vspa4ce{-0.4mm}\\
  \includegraphics[width=0.4\textwidth, clip]{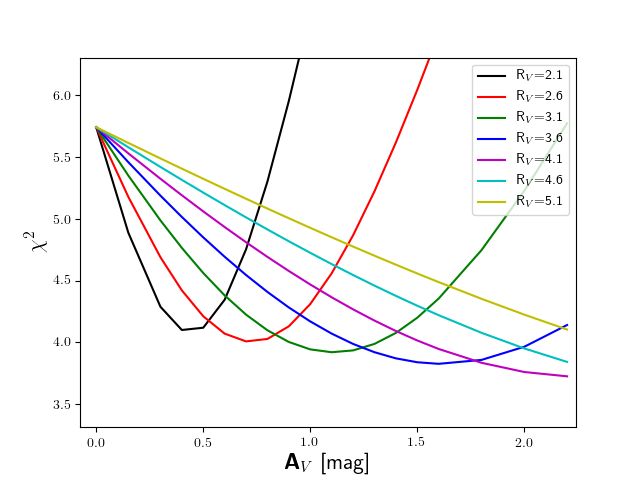}\\%\vspa4ce{-0.4mm}\\
  \caption{\label{methodi} Estimation of $A_{V}$ by $\chi^{2}$ for different $R_{V}$ as
    marked in the legend. \emph{Top:} For TW~Hya.
    \emph{Upper middle:} For SY~Cha. \emph{Lower middle:} For DK~Tau.
    \emph{Bottom:} For MY~Lup.
}
\end{center}
\end{figure}

We show our results exemplarily in Fig.~\ref{methodi} for the stars TW~Hya, DK~Tau~A, SY~Cha, and MY~Lup, which worked best using Method I in the sense, that all three progressions could be used. Since for no star a distinct $\chi^{2}$ minimum can be identified we stick to a fixed $R_{V}$=3.1. We estimate the $1\sigma$ variance in $A_{V}$ by using $\chi^{2} +1$.  For TW~Hya we do not get
$A_{V}$=0\,mag, instead for $R_{V}$=3.1 we measure an $A_{V}=1.3\pm 0.5$\,mag, which should be
clearly too high for this star, though also Method I leads to a similar value, if $R_{V}$=3.1 is chosen. Our overall lowest $\chi^{2}$ measurement corresponds to $A_{V}=0.5\pm0.2$\,mag and
$R_{V}$=2.1, which indicates a very steep extinction law but is in agreement with our Method I. The low $R_{V}$ value may be caused by
the \mbox{H$_{2}$} lines at wavelength lower than about 1450\AA\, not being
optically thin any more. %We discuss this further in Sect.~\ref{selfabsorption}.

For SY~Cha we also obtain $A_{V}=1.3\pm 0.75$\,mag for a fixed $R_{V}$=3.1, which is in agreement with Method I. Fixing R$_{V}$ to the preferred value of Method I, to 2.1,  would 
lead to identical values of $A_{V}$=0.5\,mag for Method I and Method II. 
For DK~Tau~A  we derive
$A_{V}=1.4\pm 0.9$\,mag, which is in agreement with Method I. For MY~Lup we derive $A_{V}=1.1\pm 0.7$\,mag, which is lower than the value obtained with Method I. Nevertheless, Method I yields consistent results, if we there also fix $R_{V}$=3.1.

%For $R_V$=3.1 we obtain for the two stars RECX~15 and CVSO~90 $A_V$=0.75$\pm$0.4  and 0.4$\pm$0.6\,mag, respectively. These values are in rough agreement with Method I and with the literature value $A_V$=0\,mag.

%For DM~Tau we obtain $A_V=1.8\pm0.4$\,mag for $R_V$=3.1, which clearly disagrees with Method I and the literature value of $A_V$=0\,mag. This is most probably caused by progression [0,1], which also led for Method I to much higher results, but there could easily identified as outlier and been excluded.

Unfortunately, Method II leads generally to large error bars that prevent us from identifying unique values for $A_{V}$ for the individual stars. However, the error ranges for all stars except TW~Hya overlap if one allows for the low $R_{V}$ value.

\subsection{Comparison of the two methods}
Method I has several advantages over Method II. First, while we treat all progressions together in Method II, for Method I we treat each progression separately and combine these then into one result for the A$_{V}$  of each star. Though this merging of the results comes with difficulties, it allows progressions with differing contributions to be identified. Second, if theoretical ratios lower than the observed ratios are found,
the pattern in the quotients affected often allows to identify the line, which may
contribute an erroneous flux. Since only some of these lines show also unacceptable Gaussian fits, this allows for some additional 'hand-picking' of the used lines.
We think that
the better usage of the information content of the measurements together with the better control of error introducing drawbacks of  Method I is allowing us to break
the degeneracy of $A_{V}$ and $R_{V}$ (at least partly) by comparing the different
progressions.

\end{document}